\documentclass[fleqn,usenatbib]{mnras}

\usepackage{newtxtext,newtxmath}
\usepackage[T1]{fontenc}
\usepackage{ae,aecompl}

\usepackage{graphicx}	% Including figure files
\usepackage{amsmath}	% Advanced maths commands
\newcommand{\thetaE}{\theta_{\rm E}}
\newcommand{\pirel}{\pi_{\rm rel}}
\newcommand\murel{{\mu_{\rm rel}}}
\newcommand\tE{t_{\rm E}}
\newcommand\piE{\pi_{\rm E}}
\title[VVV dark lenses]{Dark lenses through the dust: parallax microlensing events in the VVV}

\author[Z. Kaczmarek et al.]{Zofia Kaczmarek$^{1,2}$, Peter McGill$^{3}$, N. Wyn Evans$^{1}$, Leigh C. Smith$^{1}$, Łukasz Wyrzykowski$^{2}$, \newauthor Kornel Howil$^{2}$ and Maja Jab{\l}o{\'n}ska$^{2}$ \\
$^{1}$Institute of Astronomy, University of Cambridge, Madingley Rd, Cambridge CB3 0HA, UK \\
$^{2}$Warsaw University Astronomical Observatory, Al. Ujazdowskie 4, 00-478 Warszawa, Poland \\
$^{3}$Department of Astronomy and Astrophysics, University of California, Santa Cruz, CA 95064, USA}

\date{Accepted XXX. Received YYY; in original form ZZZ}

\pubyear{2022}

\begin{document}
\label{firstpage}
\pagerange{\pageref{firstpage}--\pageref{lastpage}}
\maketitle

\begin{abstract}
We use near-infrared photometry and astrometry from the VISTA Variables in the Via Lactea (VVV) survey to analyse microlensing events containing annual microlensing parallax information. 
These events are located in highly extincted and low-latitude regions of the Galactic bulge typically off-limits to optical microlensing surveys. 
We fit a catalog of $1959$ events previously found in the VVV and extract $21$ microlensing parallax candidates. The fitting is done using nested sampling to automatically characterise the multi-modal and degenerate posterior distributions of the annual microlensing parallax signal. 
We compute the probability density in lens mass-distance using the source proper motion and a Galactic model of disc and bulge deflectors. By comparing the expected flux from a main sequence lens to the baseline magnitude and blending parameter, we identify 4 candidates which have probability $> 50$ \% that the lens is dark.
The strongest candidate corresponds to
a nearby ($\approx0.78$ kpc), medium-mass ($1.46^{+1.13}_{-0.71} \ M_{\odot}$) dark remnant as lens.
In the next strongest, the lens is located at heliocentric distance $\approx5.3$ kpc. It is a dark remnant with a mass of $1.63^{+1.15}_{-0.70} \ M_{\odot}$. Both of those candidates are most likely neutron stars, though possibly high-mass white dwarfs.
The last two events may also be caused by dark remnants, though we are unable to rule out other possibilities because of limitations in the data. 

\end{abstract}


%%%%%%%%%%%%%%%%%%%%%%%%%%%%%%%%%%%%%%%%%%%%%%%%%%

%%%%%%%%%%%%%%%%% BODY OF PAPER %%%%%%%%%%%%%%%%%%

\section{Introduction}

Gravitational microlensing is a powerful technique for studying the populations of compact objects in the Milky Way. Crucially, it permits the study of bodies that emit little or no light. All that is required is the chance alignment of a foreground lens with mass $M$ and a more distance background source. In the case of perfect lens-source alignment an image of the source in a Einstein ring is formed with angular radius \citep{Chwolson1924,Einstein1936}
\begin{equation}
    \theta_{\rm E} = \sqrt{\frac{4GM}{c^{2}}\pi_{\text{rel}}},
\end{equation}
is formed. Here, $G$ is the gravitational constant, $c$ is the speed of light and $\pi_{\text{rel}}$ is the relative lens-source parallax ($\pi_{\text{lens}}-\pi_{\text{source}}$). In the case of imperfect lens-source alignment, two images of the source are formed. As the lens passes between the source and observer the source images change brightness and position giving rise to transient photometric \citep[e.g.][]{Refdal1964,Paczynski1986} and astrometric microlensing effects \citep[e.g.][]{Walker1995,DoSa,Belokurov2002}. In contrast to other methods, such as X-ray binary studies \citep[e.g.][]{Shao2020} and gravitational wave searches \citep[e.g.][]{Abbott2020}, it remains the only one sensitive to isolated stellar-mass black holes.

The all-sky averaged microlensing optical depth, or probability that a microlensing event occurs at any instant, is low $\approx 10^{-7}$ \citep[e.g.,][]{EB,Specht2020}. Therefore, microlensing searches require long-term monitoring of a large number of stars with time-series photometry. In the 1980s and 1990s, the advent of large sky surveys made it possible to propose and develop gravitational microlensing as a tool for finding dark compact objects \citep{Paczynski1986, Griest1991, Paczynski1996}. Since then, it has been applied in various searches of stellar-mass black holes~\citep[e.g.,][]{Ma02, Be02, Wyrzykowski2016}. Particularly, \cite{Wyrzykowski2020} have shown that the use of gravitational microlensing can overcome biases that other observing methods exhibit towards objects in specific evolutionary phases. Moreover, they argue that the apparent mass gap between the lightest known black holes ($\approx 5M_\odot$) and the heaviest known neutron stars of ($\approx 2 M_\odot$) can be explained with those biases. This was further explored in \citet{Mroz2021a} and \citet{Mroz2021b}, where a continuous remnant mass function derived from the analysis of OGLE-III events is presented. Finally, gravitational microlensing can be used for studying the putative population of primordial black holes~\citep{Ca16}, and exploring the viability of the remaining mass window $> 20M_{\odot}$~\citep{Gr16,Ka16, GB2018a, GB2018b}. 

While the majority of photometric microlensing events are currently detected towards the Galactic bulge, they are predominantly found by monitoring surveys in the optical wavelengths \citep[e.g.][]{Udalski2015, KMTnet2016}. This limits the search region for events to an annulus around the bulge where the interstellar extinction along the line of sight is low enough to detect events in the optical, yet the background source density is high enough for lensing to be likely \citep[e.g.][]{Wyrzykowski2015, Mroz2019}. The innermost regions of the bulge, where the microlensing event rate is high~\citep{Gould1995,EB}, are obscured by dust in the optical.  This means that populations of massive black holes expected to reside in the Galactic plane are currently off limits to optical microlensing surveys \citep{Jonker2021}. Monitoring sources in the near-infrared (NIR), however, overcomes the limitation of avoiding the Galactic plane and allows us to see through the dust.

The United Kingdom Infrared Telescope (UKIRT) microlensing survey \citep{Shvartzvald2017} first surveyed the inner regions of the bulge in the NIR. UKIRT monitored an $\approx10$ square degree patch of sky between J$2015$-J$2016$ close to the Galactic plane ($b\approx0^{\circ}$) and reported five highly extincted events missed by optical surveys. Leveraging the high event rate in the NIR \citet{Navarro2017} pioneered the use of $K_{s}-$band photometry from the VISTA Variables in the Via Lactea (VVV) survey \citep{Minniti2010} and found $182$ events in the innermost regions of the bulge. Next \citet{Navarro2018Longitude,NavarroLatitude,NavarroForesaken} extended the search for events in the VVV along a longitude strip along $b\approx0^{\circ}$ and a latitude strip along $l\approx-1^{\circ}$ finding $\approx 600$ events. \cite{NavarroFarDisk} also reported events with a source in the far disk.  Due to a search method that mainly relied on visual inspection of light curves Navarro only searched only a small section of the entire VVV footprint. Building on Navarro et. al's work, \cite{Husseiniova2021} developed a scalable machine learning algorithm to extract microlensing events over the entire VVV survey. \cite{Husseiniova2021} found $1959$ events with this method and highlighted the need for a Bayesian analysis to characterise their often sparsely sampled signals. Finally, astrometry from the VVV in combination with Gaia data \citep{GDR22018} was used to search for predictable microlensing events \citep{Refdal1964} towards the Galactic bulge \citep{McGill2019b}.

In a standard microlensing event \citep{Paczynski1986}, the only measurable parameter with physical significance is the timescale $\tE$. This is related to the lens mass $M$ through
\begin{equation}
  \tE = \frac{\sqrt{\kappa\pirel M}} {\murel} = \frac{\thetaE} {\murel}
  \label{tE}
\end{equation}
where $\kappa\approx8.14\text{mas}/M_{\sun}$ and $\pirel$ and $\murel$ are the lens-source relative parallax and proper motion respectively~\citep{Gould2000, Rybicki2018}. As there is typically no information about the parameters $\pi_{\text{rel}}$ and $\mu_{\text{rel}}$, mass measurement in a lensing event proves a difficult task. Although it is estimated that black holes should be responsible for 0.8\% of lensing events (and stellar remnants in general for 20\%) \citep{Gould2000b}, 
there are still no confirmed black hole detections from microlensing so far. A highly probable candidate has been reported very recently by \citet{Sahu2022} and \citet{Lam2022}, but its nature remains uncertain.

There are several effects observable for atypical lensing events that allow for introducing additional information tied to physical parameters of the lens \citep{Evans2003}. One such effect is the parallax deviation. The standard \citet{Paczynski1986} microlensing lightcurve assumes that the relative motions between observer, lens and source are all uniform and linear; this gives rise to a lightcurve symmetric about the peak. However, the Earth-based observer revolves around the Sun causing an acceleration effect \citep{Alcock1995}. These parallax effects break the symmetry of the lightcurve, leading to mild asymmetries~\citep[e.g.,][]{Ma02,Be02,Po05,Wyrzykowski2016} or even spectacular multiple peak behaviour~\citep[e.g.,][]{Sm02, Kruszynska2021}.

Fitting a lensing model with parallax allows us to measure the microlensing parallax vector $\boldsymbol{\pi}_{\rm E}$, defined in~\citet{Gould2004} to have a value of  
\begin{equation}
|\boldsymbol{\pi}_{\text{\rm E}}|=\piE = \sqrt{\frac{\pirel}{\kappa M}} = \frac{\pirel}{\thetaE}
\label{piE},
\end{equation}
and the direction of the relative motion of the lens with respect to the source. Measurement of the $\pi_{\text{\rm E}}$ value is a powerful way to impose additional constraints on the mass of the lens, leaving two observables and three unknowns in eqs~(\ref{tE}) and (\ref{piE}). The direction of relative motion also allows for putting additional constrains on the kinematics of the lens.

The measurement of $\thetaE$ would allow to solve the problem of determining the lens mass entirely; some possible methods are described in \citet{Lee2017}. With current observational possibilities, so far it has only been possible for several rare events -- namely, image resolution with interferometry for extremely bright events \citep{Dong2019, Cassan2021}, or astrometry for known, extremely nearby lenses \citep{Sahu2017, Zurlo2018}. In the case of astrometric lensing, the light center of images of the source gets deflected away from the position of the lens, and measurement of those deflections can be directly tied to $\thetaE$ \citep{DoSa}. \citet{Lam2020} have demonstrated that among events with the highest astrometric deviations ($\gtrsim$0.2 mas), the majority should be caused by dark remnants and especially black holes, both because of their high mass and lack of blending with lens light, which lowers the observed signal. A detection of astrometric deviation caused by an unknown, dark lens has been done for the first time, with HST imaging over six years, very recently by \citet{Sahu2022} and \citet{Lam2022} -- but there is still significant tension in the mass measurements and, in conclusion, the true nature of this object (black hole/neutron star).

The angular Einstein radius can also be derived by resolving the lens and the source years after the maximum approach (or before it - for predicted events) and measuring the relative lens-source motion, as $\thetaE = \murel \tE$ \citep[e.g.][]{Kozlowski2007, McGill2018, McGill2019}. However, it requires a very long ($\approx$10 years) coverage of observations, and is only possible for a luminous lens, which makes this method not feasible in the search for dark remnants.

Alternatively, one can use distributions of motions and distances in the Galaxy, together with the $t_{\rm E}, \vec{\pi_{\rm E}}$ measurements, to statistically infer the probability distributions of lens mass and distance. This is the approach outlined in \cite{Wyrzykowski2015} and \cite{Wyrzykowski2020}, which we adapt in this work with some modifications.

In this paper, we examine the \cite{Husseiniova2021} sample of 1959 microlensing events with an annual microlensing parallax model. For the first time, we apply a nested sampling algorithm to automatically and fully characterise the posterior distributions of parameters of the annual microlensing parallax signal. We then use this model in combination with VVV astrometry 
to identify candidate events caused by dark lenses. First, we briefly outline the reduction of VVV photometric and astrometric data used in this study. We then detail the annual parallax microlensing model and explain how we fit it to the VVV photometry. Next, we detail how we obtain an astrometric solution of the source from the VVV astrometric time-series data,
and use it together with the photometric model results to determine whether an event was caused by a dark lens building on the work of \cite{Wyrzykowski2015} and \cite{Wyrzykowski2020}. Finally, we present our analysis of a set of dark lens candidates and discuss implications for future NIR microlensing surveys of the Galactic bulge with the Roman Space Telescope (RST).

\section{VVV photometry and astrometry}

The VVV survey and its temporal and spatial extension the VVVX survey comprise nearly a decade of near-infrared observations of the Galactic bulge and southern Galactic plane. Both surveys utilise the VISTA Infrared Camera (VIRCAM) on the Visible and Infrared Survey Telescope for Astronomy (VISTA), for which \citet{sutherland15} provide more detail. Observations are initially processed by the Cambridge Astronomical Survey Unit (CASU) VISTA data flow system described by \citet{vdfs}. Our analysis is based solely on those covering the region $|l|<10,~ -10<b<5$, i.e. the region of the Galactic bulge covered by both the VVV and VVVX surveys for which we have the highest density and longest temporal baseline of observations. The selected observations were processed by the prototype VVV Infrared Astrometric Catalogue version 2 (VIRAC2, see \citealt{virac} for details of version 1) pipeline to produce calibrated astrometric and photometric time series of individual stars.

Full details of the VIRAC2 pipeline and catalogue is outside the scope of this paper, and will be provided by Smith et al. (in prep). To briefly summarise: Object detection in each observation is performed using a modified version of the DoPHOT package \citep{dophot1,dophot2}, which provides psf fitting based flux and centroid measurements; Astrometric calibration of the observations is then performed through the fitting of Chebyshev polynomials mapping array positions of stars also present in the Gaia catalogue (originally DR2 \citet{GDR22018}, more recently eDR3 \citet{GEDR32021}) to their Gaia positions propagated to the epoch of the VVV/VVVX observation using their 5-parameter Gaia astrometry; A novel, iterative star identification algorithm is then used to track detections of individual stars between epochs.

By this point in the pipeline we have a reasonably complete and clean list of stars and their calibrated astrometric time series, but their photometric time series are instrumental only. To rectify this, we employ a calibration method inspired by the "uber-calibration" approach of \citet{ubercal}. We take advantage of the overlaps in VIRCAM observations necessary to provide continuous spatial coverage to globally optimise a set of survey zero-points (one per bandpass), a set of per-bandpass per-detector per-observation photometric offsets (which take into account changes in mirror reflectivity, weather conditions, and other time-variable effects), and a per-bandpass per-detector illumination map (which takes into account local pixel scale, defective chip regions, and other highly localised effects). Since temporally variable photometric offsets are unaccounted for on scales smaller than a detector, we identified coherent structures in the offsets of individual observations of stars from their survey-average photometry, which we suspect are caused by atmospheric effects. To try to correct for these, and hence reduce their contribution to light curve scatter, we employ a further calibration stage which fits Chebyshev polynomials mapping single-epoch photometry to survey-average photometry\footnote{Details of this final photometric calibration refinement stage are available at \href{https://github.com/leigh2/coherent_residual_mapper}{github.com/leigh2/coherent\_residual\_mapper}.}.

\begin{table}
\centering
\begin{tabular}{lll} 
\hline
parameter & prior & unit \\
 \hline
 $t_{\rm E}$ & uniform(0,1000) & days \\
 $u_0$ & uniform(-2, 2) & $\thetaE$ \\
 $f_{\text{s}}$ & uniform(0,1.1) & - \\
 $t_{0}$ & uniform($t_{\text{min}}$, $t_{\text{max}}$) & days \\
 $m_{0}$ & uniform($m_{50}$ - 0.5, $m_{50}$ + 0.5) & $K_{s}$-band mag \\ 
 $\pi_{\text{EN}}$ & uniform(-3,3) & $\thetaE$ \\
 $\pi_{\text{EE}}$ & uniform(-3,3) & $\thetaE$ \\
 
\end{tabular}
\caption{Priors used in the modelling of events. $t_{\text{min}}$ and $t_{\text{max}}$ are the minimum and maximum epochs for which brightness of the event was measured in VVV, and $m_{50}$ is median magnitude during the entire coverage. Priors used for $t_{\rm E}$, $u_0$,  $f_{\text{s}}$, and $t_{0}$  $m_0$ were the same for the parallax and non-parallax models, while $\pi_{\text{EN}}$ and $\pi_{\text{EE}}$ were used only in the parallax model. Our prior for $f_{\text{s}}$ allows for small amount of negative blending which is possible due to systematics in the DoPHOT \citep{dophot1,dophot2} processing pipeline \citep{Smith2007} used in the VVV data reduction.}
\label{table:priors}
\end{table}

\section{Methods}

\subsection{Light curve modelling}
\label{sec:LCmodelling}

In this paper, we fit two different point source, point lens microlensing models to the light curve data. The first assumes the standard \cite{Paczynski1986} rectilinear lens-source trajectory parameterisation 
\begin{equation}
    \boldsymbol{u}_{\text{lin}}(t) = \boldsymbol{u_{0}} + \frac{t-t_{0}}{t_{\rm E}}\boldsymbol{
    \hat{\mu}}_{\text{rel}}.
    \label{linear_traj}
\end{equation}
Here, $\boldsymbol{u_{0}}$ is the normalised lens-source angular separation vector with direction towards the source position, $t_{0}$ is the time of lens-source closest approach, $t_{\rm E}$ is the Einstein timescale, and $\boldsymbol{\hat{\mu}}_{\text{rel}}$ is the unit vector in the direction of the relative lens-source proper motion. The second is an annual parallax model, which additionally takes into account the observer's acceleration around the Sun. We use the \cite{Gould2004} geocentric conventions and parameterisation. In this, the lens-source normalised separation is given by the sum of $\boldsymbol{u}_{\text{lin}}$ and an offset term, 
\begin{equation}
    \boldsymbol{u}_{\text{par}}(t) = \boldsymbol{u}_{\text{lin}}(t) + \boldsymbol{\pi}(t; \boldsymbol{\pi}_{\rm E}).
    \label{par_traj}
\end{equation}
Here, $\boldsymbol{\pi}_{\text{\rm E}}= \pi_{\text{EE}}\hat{\boldsymbol{\rm E}} + \pi_{\text{EN}}\hat{\boldsymbol{n}}$ is the microlensing parallax vector with components $\pi_{\text{EE}}$ and $\pi_{\text{EN}}$ in the local north ($\boldsymbol{\hat{n}}$) and east ($\boldsymbol{\hat{\rm E}}$) directions. The Sun's positional offset terms needed for $\boldsymbol{\pi}(t; \boldsymbol{\pi_{\rm E}})$'s computation \citep[detailed in][]{Gould2004} were retrieved using the astropy PYTHON package \citep{Astropy2013, Astropy2018}, which uses values computed from NASA JPL’s Horizons Ephemeris\footnote{\url{https://ssd.jpl.nasa.gov/horizons/}}. We expanded the \cite{Gould2004} parallax trajectory approximation around the posterior median $t_{0}$ of the previous rectilinear fit obtained in \cite{Husseiniova2021} for each event. 

For a given trajectory model $\boldsymbol{u}_{\mathcal{M}}(t)$ and under the point source, point lens assumption the amplification of the unresolved source images is \citep[e.g.][]{Paczynski1986},
\begin{equation}
    A_{\mathcal{M}}(t) = \frac{u_{\mathcal{M}}^{2}(t)+2}{u_{\mathcal{M}}(t)\sqrt{u_{\mathcal{M}}^{2}(t)+4}}.
    \label{pspl_amp}
\end{equation}
Here $u_{\mathcal{M}}^{2}(t)=|\boldsymbol{u}_{\mathcal{M}}(t)|$ is the magnitude of the normalised lens-source separation vector. Assuming some non-zero amount of blended light not from the source (from the lens or otherwise), the observed magnitude of the lens-source blend is 
\begin{equation}
    m_{\mathcal{M}}(t;\boldsymbol{\theta}) = m_{0} - 2.5\log_{10}\left[f_{\text{s}}A_{\mathcal{M}}(t) + (1 - f_{\text{s}})\right].
\end{equation}
where the blending parameter $f_{\text{s}}$ is the fraction of light that is contributed by the source, and $\boldsymbol{\theta}$ is the vector of model parameters. The standard rectilinear model has five parameters, namely $\boldsymbol{\theta}=[t_{\text{\rm E}}, u_{0}, t_{0}, m_{0}, f_{\text{s}}]$. The inclusion of parallax gives a seven parameter model $\boldsymbol{\theta}=[t_{\text{\rm E}}, u_{0}, t_{0}, m_{0}, f_{\text{s}}, \pi_{\text{EE}}, \pi_{\text{EN}}]$. Here, $u_{0}=|\boldsymbol{u_{0}}|$ is the magnitude of the minimum lens-source normalised separation vector. 

Let $D_{i}=\{t_{i},m_{i}, \sigma_{m_{i}}\}$ be a data point, comprising a time of observation $t_{i}$, observed magnitude $m_{i}$, and magnitude error bar $\sigma_{m_{i}}$. Let a light curve for a given event be denoted as $\mathcal{D}=\{ D_{i} \}_{i=1}^{n}$ and be the set of $n$ data points. Under the assumption that $t_{i}$ is known perfectly for each $D_{i}$ and the data are scattered with independent Gaussian noise with variance $\sigma^{2}_{m_{i}}$, the log likelihood for a light curve given a particular model ${\mathcal{M}}$ is,  
\begin{equation}
    \ln p(\mathcal{D}|\boldsymbol{\theta},\mathcal{M}) = -\frac{1}{2}\boldsymbol{r}^{T}\textbf{\textsf{K}}^{-1}\boldsymbol{r}-\frac{1}{2}|\textbf{\textsf{K}}| - \frac{n}{2}\ln2\pi.
    \label{eq:log_like}
\end{equation}
Here $\textbf{\textsf{K}}=\text{diag}(\sigma^{2}_{m_{1}},..,\sigma^{2}_{m_{n}})$ is an $n\times n$ diagonal covariance matrix, and $\boldsymbol{r} = \left[m_{1}-m_{\mathcal{M}}(t_{1};\boldsymbol{\theta}),...,m_{n}-m_{\mathcal{M}}(t_{n};\boldsymbol{\theta})\right]^T$ is a vector of length $n$ and is the residual between the model and data.

With the likelihood for the two models in hand, the posterior distribution of the model parameters is given by Bayes theorem as
\begin{equation}
    p({\boldsymbol\theta}|\mathcal{D}, \mathcal{M}) = \frac{p(\mathcal{D} |{\boldsymbol\theta}, \mathcal{M})p(\boldsymbol{\theta}| \mathcal{M})}{p(\mathcal{D}|\mathcal{M})}
    \label{eq:bayes}
\end{equation}
Here, $p(\boldsymbol\theta| \mathcal{M})$ is the prior distribution of the model parameters, while
\begin{equation}
    p(\mathcal{D}|\mathcal{M}) = \int_{\Omega_{\boldsymbol{\theta}}}
    p(\mathcal{D}|\boldsymbol{\theta},\mathcal{M})p(\boldsymbol{\theta}| \mathcal{M})d\boldsymbol{\theta},
    \label{eq:evidence}
\end{equation}
is the Bayesian evidence, or the probability of the data given the model, while $\Omega_{\boldsymbol{\theta}}$ is the space of all possible parameter values. In this study, our strategy for both models is to choose weakly informative priors for all parameters that constrain their values to reasonable areas of the parameter space. The priors are listed in Table \ref{table:priors}. The prior factorises over all the model parameter for both models.

To characterize the parameter posterior distributions, we use nested sampling. We choose this sampling method motivated by its advantage over traditional Monte Carlo Markov Chain (MCMC) approaches \citep[e.g.][]{Foreman-Mackey2013} when handling multi-modal distributions. This is important for microlensing events found in the VVV because of the typically sparse sampling of the microlensing signals, which can result in multi-modality for microlensing parameters even in the rectilinear trajectory model \citep[see e.g. Figure 12 of ][]{Husseiniova2021}. Crucially, nested sampling also allows the characterisation of the (at least two-fold) degeneracy in the parallax trajectory model. With the addition of the parallax deviation, the positive and negative $u_{0}$ solutions are no longer completely indistinguishable but yield a pair of related solutions. Physically, we do not know on which side the source passes the lens, so positive $+u_{0}$ and negative $-u_{0}$ solutions arise, yielding roughly opposite $\vec{\pi_{\rm E}}$ directions~\citep{Sm03}. Moreover, some events are subject to further jerk-parallax degeneracies and have four different solutions, as described in detail in \cite{Gould2004}.

In previous analyses of microlensing parallax events, MCMC has been used to characterise the parameter posterior distributions \citep[e.g.][]{Wyrzykowski2016, Wyrzykowski2020,Golovich2020}. 
For example, the parallax lensing event Gaia18cbf has been recently analysed by \cite{Kruszynska2021}, who found three possible solutions using \textit{Gaia} data alone, and two possible solutions with follow-up. In most studies \citep{Wyrzykowski2016, Wyrzykowski2020, Kruszynska2021}, the modes of the posterior are sampled and analysed separately, yielding alternative solutions for lens mass and distance. A possible disadvantage is not returning the relative probability of those solutions and only using $\chi^2$ as an indicator of weak preference towards one of the solutions. Additionally, human supervision is needed to set relative boundaries between MCMC solutions, which requires significant time and effort to analyse larger candidate samples. Alternatively, \cite{Golovich2020} chose to only sample the $u_0>0$ solutions.

In contrast to previous work, nested sampling has the advantage of allowing us to characterise all posterior modes and propagate all uncertainty downstream in our analysis. A proof of concept in application to a binary microlensing event can be seen in \cite{Sharan2019}. Without the need to apply arbitrary cuts in order to separate solutions, we are able to fully automate the modelling and greatly reduce the time and effort needed to provide results for parallax and non-parallax models for all sources, while still sampling the entire parameter space. Given the relatively large size of the sample (1959 events), this is a significant improvement.

Finally, nested sampling also has the advantage of being able to compute an estimate for the Bayesian evidence in eq.~(\ref{eq:evidence}). This quantity provides a metric that we use to compare the fit of the linear and parallax trajectory models, which naturally penalises the increased complexity and flexibility of the parallax trajectory model to fit the data.  We fit all $1959$ events from the \cite{Husseiniova2021} sample to the linear and parallax trajectory model using the dynamic nested sampling algorithm \citep{Higson2019} implemented by \citet[][the {\tt dynesty} code]{DYNESTY}. We use random walk sampling \citep{Skilling2006} with multiple bounding ellipsoids, and with $1\,000$ initial live points. We adopt a stopping criteria in the remaining fractional evidence of $0.01$, and allocate $100$ per cent of weight on computing the posterior distributions. For the final 4 selected candidates, to ensure good accuracy in the following lens mass-distance inference, we perform a re-run with higher resolution, with $5\, 000$ initial live points and a stopping criterion in the remaining fractional evidence of $0.001$.

\subsection{Dark lens probability \label{DarkLensCode}}

To estimate the lens mass and distance, we need to compute the posterior distribution of microlensing parameters, given the lightcurve and the proper motion of the source. We employed the method described in \cite{Wyrzykowski2016} and \cite{Mroz2021a}, which we briefly summarize here. The underlying Galactic model has both a bulge and disc lens population. The deflectors can lie in a double exponential thin or thick discs or in the bulge itself. The intrinsic velocity distributions are all Gaussians with fixed dispersions and means. The mass function of the lenses is a power-law.

In \cite{Mroz2021a}, the source distance is fixed to 8 kpc. Here, we allow a distribution of source distances. This is motivated by the nature of our near-infrared data, which  allows us to partially see through the central regions of the Galaxy. The new modification allows for the possibility of source stars belonging to different stellar populations, either in the Galactic bulge or outside it in the disc. 

The {\tt Dark Lens} code assumes that the entire blended flux is being contributed by the lens. We do not consider the scenario where a third (background) star is also partially contributing to the observed flux. This assumption imposes stronger cuts on the prospective dark remnant candidates, and results in a lower estimate on the dark lens probability. As we do not know the distance to the lensing objects, which can lie anywhere in the Galaxy along the line of sight to the source, there is some uncertainty about the impact of extinction on the flux from the lens. We constrain the probabilities assuming this extinction to lie between zero and the extinction at the distance of 8 kpc in the source direction. This gives an upper and lower limit respectively to the probability of the lens being a dark remnant. For the lower limit, we use the extinction maps for the VVV to obtain values appropriate for the K$_s$-band for the healpix containing the candidates.

For each event, the code uses the microlensing model parameters with their scatter, the proper motion of the source and assumed distance of the source in order to derive the probability density for mass and distance of the lens. For each combination of parameters, we obtain a probability density in the mass-distance space. Moreover, for each combination of the lens mass, distance and blending parameter, we compute the expected brightness of a Main Sequence star and compare it with the observed amount of light computed from the baseline magnitude and the blending parameter. The integral of the probability density for the dark lens divided by the total integral of the probability density in the blending space defines the dark remnant probability for an event.

\begin{figure}
	\includegraphics[width=\columnwidth]{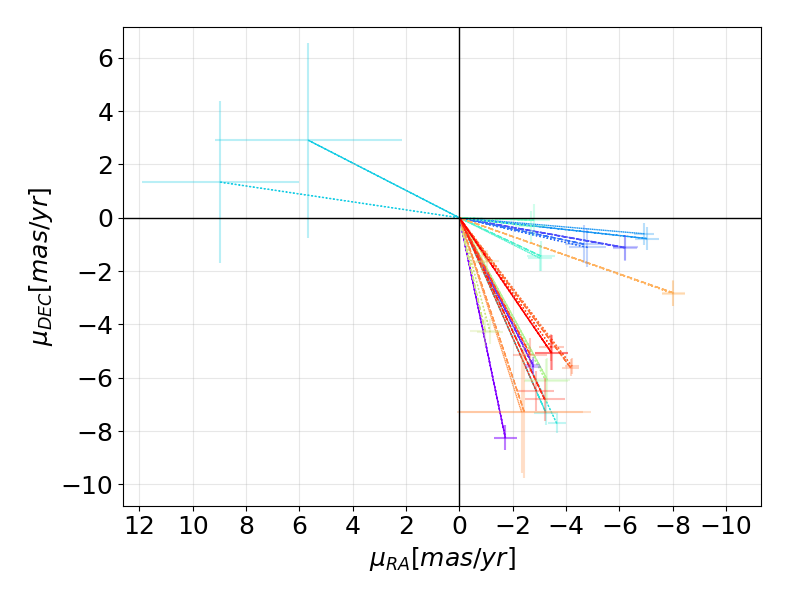}
	\includegraphics[width=\columnwidth]{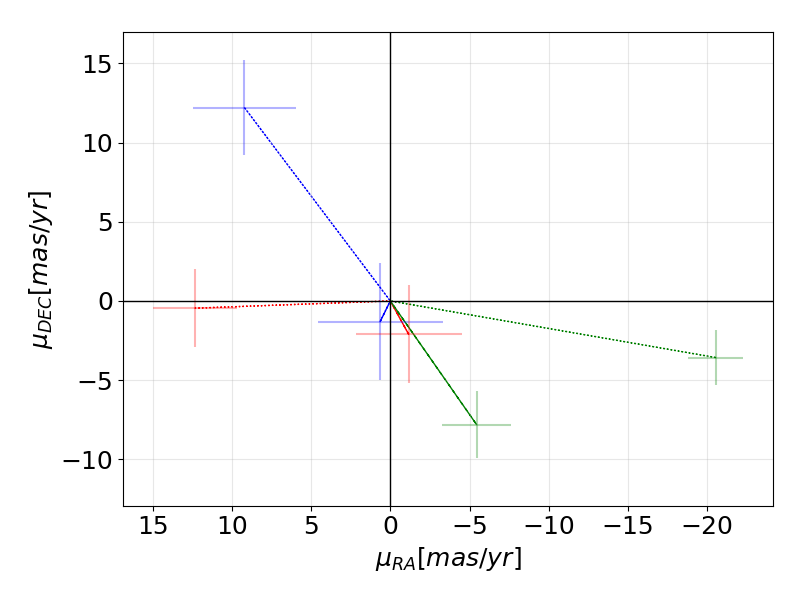}
    \caption{Upper: Masked (dashed) and unmasked (dotted) fits of 2D proper motions for 18 sources for which the proper motions are not changed significantly between fits (are within 1$\sigma$ of each other). Each colour corresponds to a different source. Lower: The same, but for the 3 sources for which the proper motions do change significantly between fits.}
    \label{fig:unchanged_fits}
\end{figure}
 
\begin{figure*}
	\includegraphics[width=1.8\columnwidth]{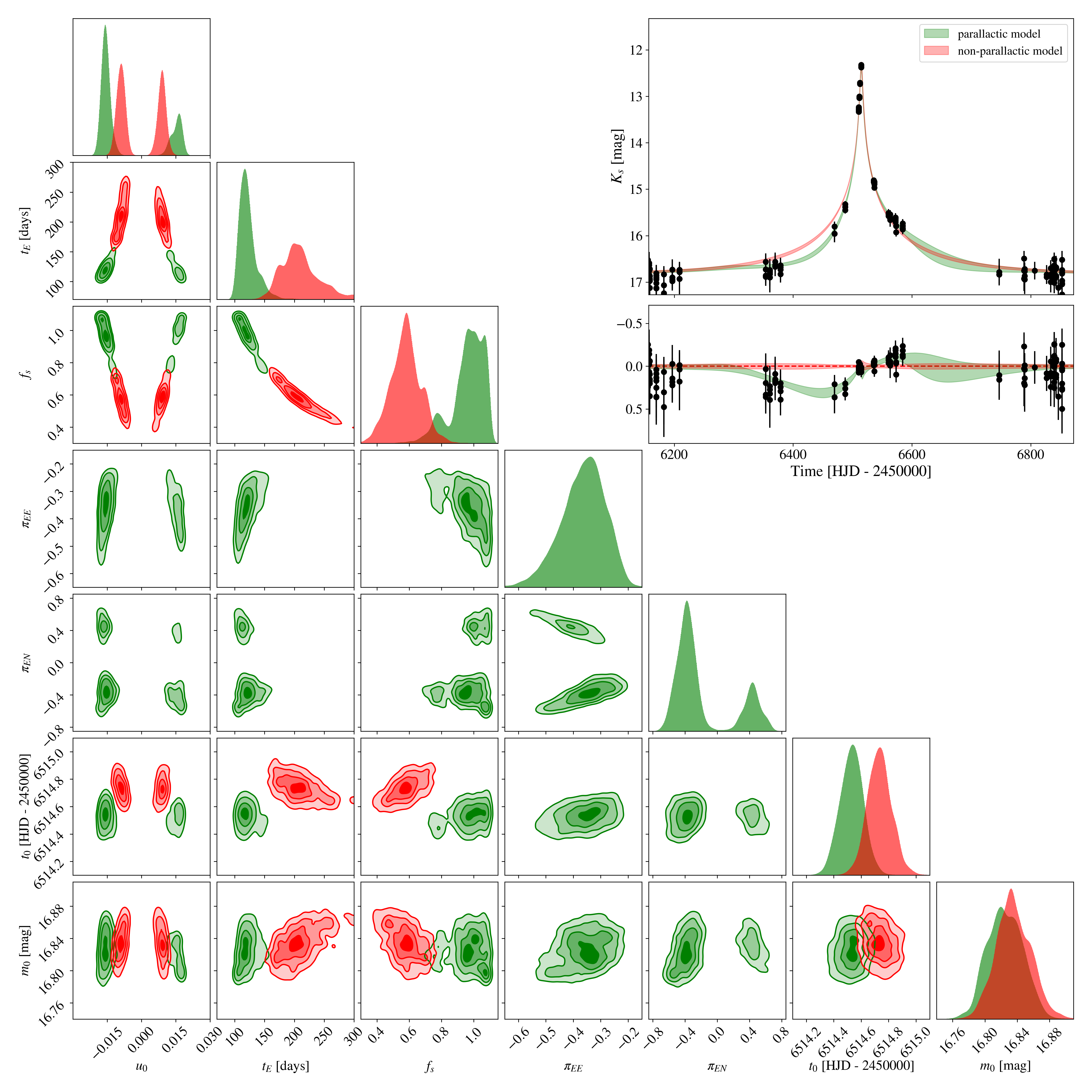}
    \caption{Main figure: Corner plot of posterior distributions of lensing parameters for the parallax and non-parallax models for event VVV-2013-BLG-0324. Top right: Lightcurve of the lensing event VVV-2013-BLG-0324. Boundaries of shaded regions represent the 10th and 90th percentiles for the non-parallax and parallax models. Residuals are plotted with respect to the median of the non-parallax model.}
    \label{fig:event1_corner_LC}
\end{figure*}
\begin{figure*}
	\includegraphics[width=2\columnwidth]{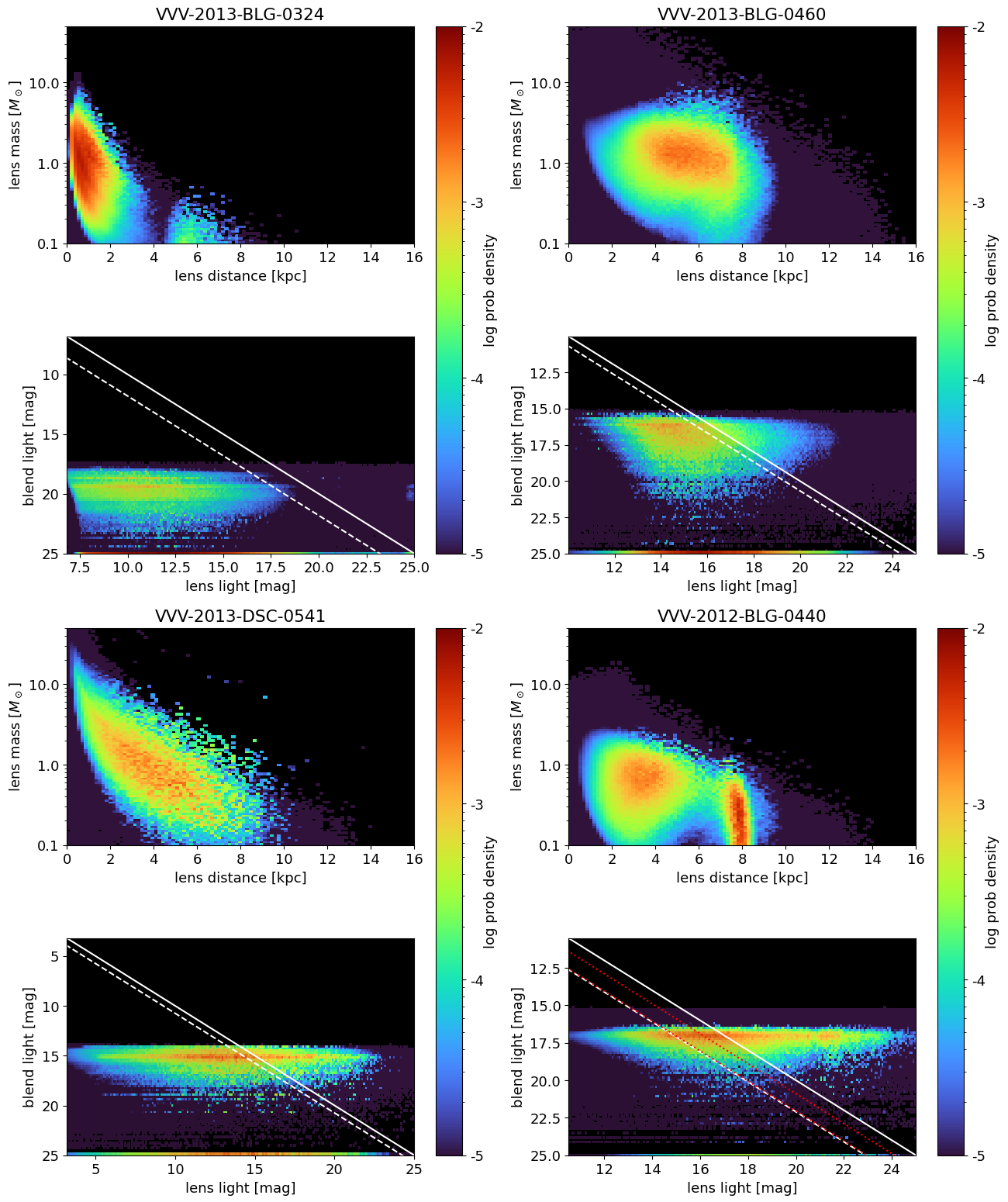}
    \caption{Analysis of the possible physical parameters of the lens in all events. Each subplot consists of a probability density plot for the mass and distance of the lens (top) and a comparison of the light expected from a luminous main sequence lens located at this mass and distance and the blend light observed in the event (bottom). The white lines mark the boundary between a dark remnant and a main sequence star in case of no extinction (solid) and extinction equal to that at 8 kpc (dashed). The subplot for VVV-2012-BLG-0440 also includes two red dotted lines marking that boundary for extinction at the median lens distance for each (disk and bulge) solution.}
    \label{fig:mass_distance}
\end{figure*}

\section{The Candidates}

\subsection{Events with strong parallax signal}
\label{initial_selection}

For each of the 1959 microlensing events in \citet{Husseiniova2021}, we calculate the Bayes factor:
\begin{align}
K = \frac{p(\mathcal{D}|\mathcal{M}_{\rm par})}{p(\mathcal{D}|\mathcal{M}_{\rm lin})}
\end{align}
to select those for which the parallax model is strongly preferred. Here, $M_{\rm par}$ denotes the parallax model and $M_{\rm lin}$ the non-parallax model. To obtain $P(D|M)$, we use the log-likelihoods directly provided within the nested sampling results from {\tt dynesty}. After applying the cut of $K>100$, we are left with 176 events.

The lightcurves of these events, together with their fits provided by both non-parallax and parallax modelling, were then inspected visually by three of the authors. They were scored as -1 (no), 0 (unsure) or 1 (yes). Events typically voted as -1 contained incomplete data, where the model usually contained very sharp brightness changes in gaps between datapoints that were impossible to verify. Events voted as 0 were considered consistent with the parallax model but still had low cadence of observations, gaps in critical regions of the lightcurve or very high scatter, making it hard to constrain the model. Events voted as 1 were the ones with a well-defined parallax lensing model for the lightcurve. With a threshold of total score $\geq$ 0, we select the final set of 21 events for further analysis. Though the rejected subset might still contain interesting events, it is impossible to obtain useful information about them due to the limitations of the dataset.

We remark that selecting events with clear parallax signal does introduce a bias in favour of nearby and low-mass lenses. \citet{Lam2020} demonstrated that events caused by BHs naturally reside within the large $t_{\rm E}$ and low $\pi_{\rm E}$ region of the parameter space, and proposed a selection criterion of $t_{\rm E} > 120$ days and $\pi_{\rm E} < 0.08$. A non-detection of parallax effect in the lightcurve can therefore also be used as an indicator of a high lens mass \citep[e.g.][]{Karolinski2020}. However, without the additional information from accurately measuring ${\boldsymbol \pi}_{\rm E}$, it is difficult to constrain the lens mass-distance distribution and prospects for determining the nature of the lens are limited. 

\subsection{VVV astrometric solution \label{VIRAC2astrometry}}

Here, we describe how to extract the positions and proper motions of the source from the VVV astrometry for our 21 candidates. We obtain astrometric time-series 2D data points for candidates from the preliminary VIRAC2 data. It is not feasible to use \textit{Gaia} data, as it is very incomplete in the heavily obscured regions towards the Galactic bulge covered by our sample.

The standard pipeline performs 5-parameter ($\alpha_0, \delta_0, \mu_{\alpha*}, \mu_{\delta}, \pi$) fits to the time-series positional data. Our main requirement is a reasonable measurement of the source proper motion ($\mu_{\alpha*}, \mu_{\delta}$). However, the true parallax of the stars in our sample is likely to be significantly smaller than the astrometric uncertainties. To avoid overfitting to noise, we refit the time series astrometry using a 4-parameter ($\alpha_0, \delta_0, \mu_{\alpha*}, \mu_{\delta}$) straight line fit. Both the high source densities in Galactic plane fields in the near-infrared and the variable nature of ground-based observation quality can impede associations of individual detections with unique stars. To cope with erroneous data points, the astrometry fitting algorithm measures residuals using 5-fold cross validation, rejects outlying data points at the level of 5$\sigma$, and then refits using all remaining data points.

Care is taken to avoid the impact of astrometric lensing on our fitting of the proper motions of the sources, although such signals will generally be subtle. Our main point of concern is that the data points at the maximum approach of the lens and the source, where the motion from a changing astrometric deflection is the most influential, are amplified and therefore might dominate the proper motion due to their generally improved astrometric precision. To investigate this, we compare `masked' and `unmasked' fits. The masked fits are obtained having rejecting all data points that had epochs corresponding to $u <3$ (the $u$ value used in this criterion was the median of $u$ values obtained from the parallax model samples). The unmasked fits uses all data points available. As shown in Fig.~\ref{fig:unchanged_fits}, the majority of sources in our sample have similar results for the masked and unmasked fits. However, 3 candidates show significant (>1$\sigma$) differences between the two.  

The changing astrometric solutions are not in themselves an indication of a lensing signal. The 3 sources with changed solutions all have baseline magnitudes $K_{\rm s} > 16$. By contrast, only 2 of the 21 sources have a baseline $K_{\rm s}$ band magnitude above 16 and exhibited no significant change between astrometric solutions. Moreover, the error bars on the 'masked' fits for those sources are large. The change in fits could be either caused by lensing effects or by the high errors on the low-magnitude points remaining after the masking procedure, making it impossible to accurately fit the proper motions. Another potential explanation is the effect of blending with the lens or background sources. In case of significant blending, the observed positions and proper motions are a weighted (by fluxes) average of positions and proper motions of two or more luminous objects. However, when the source is amplified, the weights of this average change; at that time, the observed proper motion is closer to that of the source. Two of the three events in this group have very low blending parameters ($f_{\text{s}} = 0.08^{+0.04}_{-0.03}$ and $f_{\text{s}} = 0.20^{+0.16}_{-0.07}$), which is consistent with this hypothesis.

If the assumption that all light is coming from either the source or the luminous lens were true, this would allow for using time-series astrometry to obtain a simultaneous fit of proper motions of the source and the lens. Using the measured relative proper motion $\mu_{\rm rel}$ in combination with $\tE$ obtained from the lightcurve would give $\thetaE$ and completely solve the event, including the lens mass. However, all 3 events are situated in very crowded regions, close to the Galactic Centre. To carry out such an analysis, we must identify and account for all the objects influencing the observed position, including the lens and the background stars. This is in principle possible with high-resolution imaging.

In the end, we decided to use the unmasked, non-parallax fits for all events as input into the {\tt Dark Lens} code described in \ref{DarkLensCode}. To ensure that we are still including the possibility of astrometric signal appearing, we tried simultaneous fitting of astrometry and photometry for those 3 objects. The attempts at simultaneous fitting did not lead to conclusive results with the low accuracy of VVV astrometric data being the limiting factor. Additionally, as shown recently by \citet{Sahu2022} and \citet{Lam2022}, even in case of highly accurate data there are very significant tensions between fit results favouring astrometric and photometric data. This leaves the necessity to conduct follow-up observations to completely exclude the possibility of impact of astrometric signal on the fits. 

\subsection{High dark lens probability events}

With the astrometry in hand, we provide the outputs from the {\tt Dark Lens} code -- the mass, distance and dark lens probabilities -- for all 21 candidates in Table~\ref{table:allcandidates}. This table is also available in the online supplementary material. Additionally, we provide full astrometry and photometry for those events in Table \ref{table:samplephotometry} which is available in the online supplementary material.
We find 8 candidates for which the probability of the lens (upper limit, or the case of zero extinction) being a dark remnant was calculated to be $>50\%$. None of those events have been identified by OGLE, and only one (VVV-2012-BLG-0440) occurs in the \citet{Navarro2017} sample. 
We then examined each candidate individually. For events with low $f_{\text{s}}$, the recovered astrometric solutions are not reliable. The inferred proper motions of the source are then heavily influenced by the proper motion of the blend (the lens or a background star).

For this reason, we rejected 4 of those 8 dark remnant candidates: VVV-2015-BLG-0149, VVV-2012-BLG-0245, VVV-2012-BLG-0176, VVV-2012-BLG-0255. For the first two, the blending parameter values are rather well-constrained and very low, indicating that a large majority of the flux is contributed by the blend ($f_{\text{s}} = 0.08^{+0.04}_{-0.03}$ and $f_{\text{s}} = 0.20^{+0.16}_{-0.07}$ respectively). For the latter two, the distributions of $f_{\text{s}}$ are very wide and flat, spanning the entire parameter space of physically possible solutions ($f_{\text{s}} = 0.66^{+0.27}_{-0.32}$ and $f_{\text{s}} = 0.67^{+0.29}_{-0.33}$ respectively).

We limit the final, high-probability dark remnant candidates to the remaining 4 events. We re-run the modelling for those candidates with higher resolution settings. All of them have $f_{\text{bl,median}} > 0.8$. One (VVV-2012-BLG-0440) has a bi-modal Gaussian-like distribution of the blending parameter. The remaining 3 all have a distribution of $f_{\text{s}}$ that is well-constrained and consistent with 1.

\begingroup
\renewcommand{\arraystretch}{1.4}
\begin{table*}
\caption{Results of running the analysis described in section \ref{DarkLensCode} on 21 selected events. Inferences on $f_{\text{s}}$, the lens mass, lens distance, and dark lens probability are shown. Median values with $84$th-$50$th percentile indicated as a superscript and $16$th-$50$th percentile indicated as a subscript are reported. The lower and upper bounds on the dark lens probability correspond to assumed extinction to the lens equal to that at 8 kpc for the position of the event on the sky and no extinction to the lens, respectively. Right ascension (RA) and declination (DEC) are from results of VIRAC \citep{virac} version $2$. Positions are on the International Celestial Reference Frame at epoch $2015.5$ Julian years, and were calculated using reference stars from Gaia Data Release 2. The $f_{\text{s}}$ inference from modelling the lightcurve is reported to indicate which solutions should be treated with caution because of high blending.
Events highlighted in bold are analysed in detail, and values shown for them correspond to high-resolution re-runs (see last paragraph in Subsection \ref{sec:LCmodelling}).} 
\begin{tabular}{|l|c|c|c|c|c|c|}
\hline
Event ID & \multicolumn{1}{l|}{RA [deg] } & \multicolumn{1}{l|}{DEC [deg]} & \multicolumn{1}{l|}{$f_{\text{s}}$} & \multicolumn{1}{l|}{$M_{L}$ [$M_{\odot}$]} & \multicolumn{1}{l|}{$D_{L}$ [kpc]} & \multicolumn{1}{l|}{Dark lens probability [lower-upper]} \\ \hline
VVV-2012-BLG-0245 & $271.0681752$ & $-19.3583881$ & $0.20^{+0.14}_{-0.07}$ & $0.50^{+0.28}_{-0.20}$ & $1.49^{+0.54}_{-0.50}$ & $0.940-0.947$ \\
VVV-2012-BLG-0255 & $269.0084262$ & $-21.994012$ & $0.67^{+0.29}_{-0.33}$ & $0.08^{+0.17}_{-0.06}$ & $0.65^{+0.96}_{-0.35}$ & $0.644-0.675$ \\ 
VVV-2014-BLG-0227 & $271.5624003$ & $-24.4531952$ & $0.08^{+0.05}_{-0.02}$ & $0.16^{+0.65}_{-0.13}$ & $0.96^{+0.62}_{-0.48}$ & $0.452-0.477$ \\
VVV-2012-BLG-0615 & $265.1067589$ & $-24.7612246$ & $0.89^{+0.14}_{-0.16}$ & $0.25^{+0.25}_{-0.12}$ & $3.30^{+1.19}_{-1.05}$ & $0.179-0.186$ \\ 
\textbf{VVV-2013-BLG-0460} & $\mathbf{269.9306299}$ & $\mathbf{-25.4220515}$ & $\mathbf{0.91^{+0.12}_{-0.21}}$ & $\mathbf{1.63^{+1.15}_{-0.70}}$ & $\mathbf{5.26^{+1.46}_{-1.36}}$ & $\mathbf{0.857-0.912}$ \\
\textbf{VVV-2013-BLG-0324} & $\mathbf{267.6871261}$ & $\mathbf{-26.8092714}$ & $\mathbf{0.98^{+0.08}_{-0.11}}$ & $\mathbf{1.46^{+1.13}_{-0.71}}$  & $\mathbf{0.78^{+0.51}_{-0.35}}$ & $\mathbf{0.991-0.995}$ \\ 
VVV-2015-BLG-0149 & $269.7239395$ & $-27.8991351$ & $0.08^{+0.04}_{-0.03}$ & $0.14^{+0.21}_{-0.08}$ & $0.30^{+0.18}_{-0.13}$ & $0.477-0.511$ \\
VVV-2013-BLG-0114 & $265.6348722$ & $-29.5219913$ & $0.73^{+0.25}_{-0.32}$ & $0.05^{+0.08}_{-0.03}$ & $1.15^{+0.77}_{-0.60}$ & $0.258-0.304$ \\
VVV-2012-BLG-0543 & $263.7554844$ & $-30.0823798$ & $0.69^{+0.28}_{-0.30}$ & $0.07^{+0.10}_{-0.04}$ & $1.91^{+1.19}_{-0.93}$ & $0.034-0.038$ \\
VVV-2012-BLG-0570 & $263.8670587$ & $-31.8542696$ & $0.46^{+0.30}_{-0.20}$ & $0.03^{+0.07}_{-0.02}$ & $2.86^{+1.83}_{-1.60}$ & $0.024-0.037$ \\
VVV-2013-BLG-0452 & $270.9247559$ & $-32.2202497$ & $0.15^{+0.04}_{-0.04}$ & $0.03^{+0.05}_{-0.02}$ & $0.53^{+0.42}_{-0.29}$ & $0.058-0.061$ \\
VVV-2013-BLG-0423 & $267.6273383$ & $-32.2718324$ & $0.84^{+0.16}_{-0.20}$ & $0.06^{+0.08}_{-0.03}$ & $0.85^{+0.76}_{-0.47}$ & $0.213-0.245$ \\ 
VVV-2012-BLG-0472 & $263.4082397$ & $-33.4860459$ & $0.24^{+0.18}_{-0.11}$ & $0.01^{+0.02}_{-0.01}$ & $0.98^{+0.83}_{-0.50}$ & $0.006-0.008$ \\ 
\textbf{VVV-2012-BLG-0440} & $\mathbf{262.0432834}$ & $\mathbf{-34.580735}$ & $\mathbf{0.81^{+0.09}_{-0.31}}$ & $\mathbf{0.73^{+0.52}_{-0.39}}$ & $\mathbf{3.70^{+3.69}_{-1.19}}$ & $\mathbf{0.216-0.555}$ \\ 
VVV-2012-BLG-0176 & $269.0632744$ & $-34.6781268$ & $0.66^{+0.27}_{-0.32}$ & $0.12^{+0.20}_{-0.09}$ & $1.07^{+0.63}_{-0.52}$ & $0.578-0.584$ \\
VVV-2013-BLG-0370 & $260.1191353$ & $-36.0826258$ & $0.76^{+0.22}_{-0.30}$ & $0.42^{+0.40}_{-0.28}$ & $3.52^{+2.34}_{-1.47}$ & $0.025-0.077$ \\
VVV-2013-DSC-0437 & $259.6928235$ & $-39.3839553$ & $0.46^{+0.31}_{-0.21}$ & $0.02^{+0.02}_{-0.01}$ & $0.84^{+0.57}_{-0.45}$ & $0.059-0.067$ \\
VVV-2015-DSC-0007 & $247.6672882$ & $-45.6782516$ & $0.62^{+0.29}_{-0.30}$ & $0.04^{+0.06}_{-0.03}$ & $1.75^{+1.16}_{-0.71}$ & $0.006-0.008$ \\ 
VVV-2013-DSC-0136 & $234.5936454$ & $-56.4009834$ & $0.69^{+0.25}_{-0.24}$ & $0.02^{+0.03}_{-0.01}$ & $1.23^{+0.98}_{-0.69}$ & $0.041-0.048$ \\
VVV-2013-DSC-0135 & $225.7477887$ & $-60.3183439$ & $0.85^{+0.17}_{-0.26}$ & $0.02^{+0.03}_{-0.01}$ & $1.50^{+1.54}_{-0.84}$ & $0.283-0.291$ \\
\textbf{VVV-2013-DSC-0541} & $\mathbf{193.5259103}$ & $\mathbf{-62.1339938}$ & $\mathbf{0.85^{+0.17}_{-0.21}}$ & $\mathbf{2.07^{+3.60}_{-1.27}}$ & $\mathbf{2.80^{+2.05}_{-1.52}}$ & $\mathbf{0.689-0.751}$ \\
\hline
\end{tabular}
\label{table:allcandidates}
\end{table*}
\endgroup

\subsection{Best candidates}

\begin{figure*}
	\includegraphics[width=1.8\columnwidth]{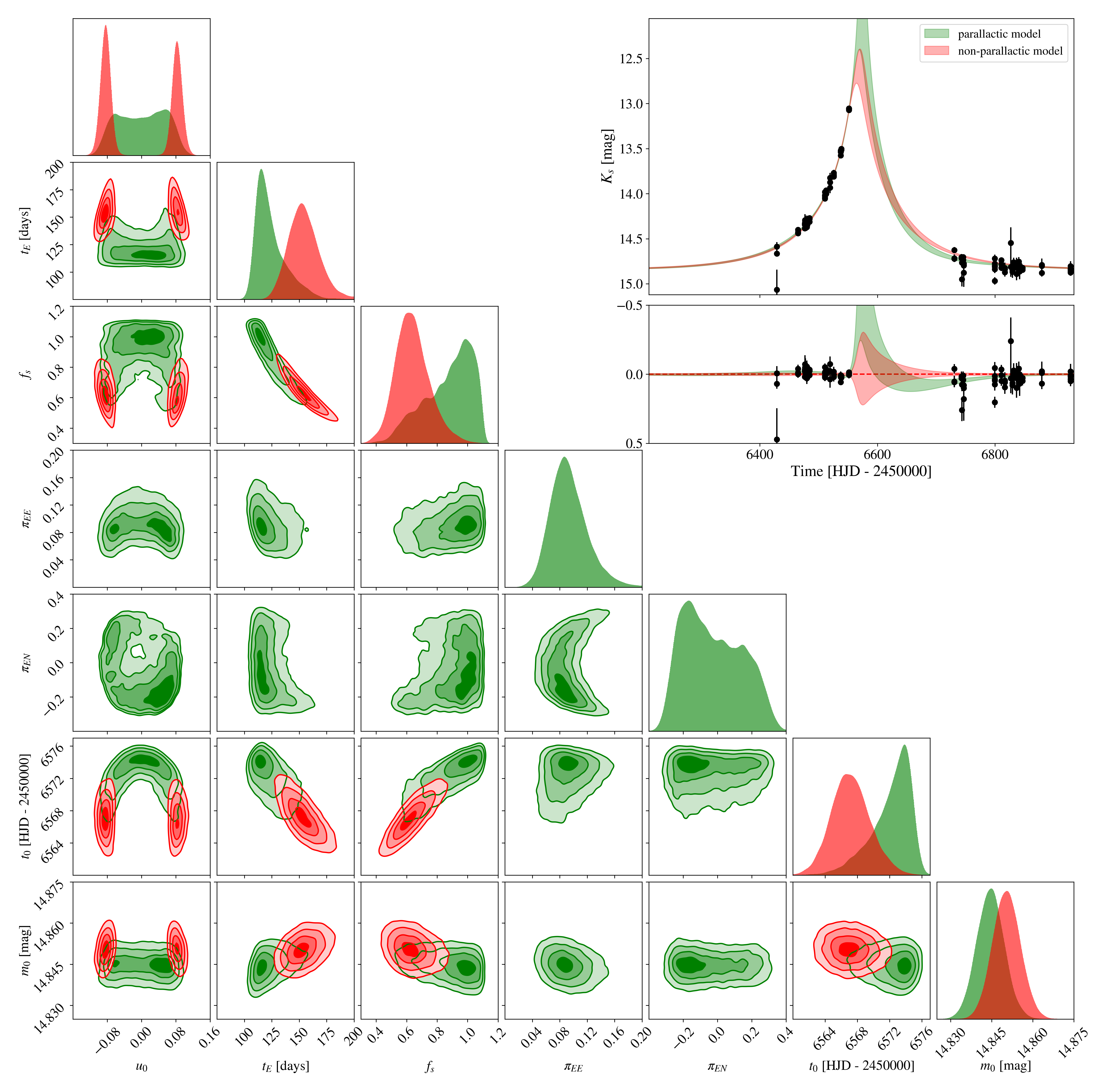}
    \caption{As Fig.~\ref{fig:event1_corner_LC}, but for event VVV-2013-BLG-0460.}
    \label{fig:event2_corner_LC}
\end{figure*}
\begin{figure*}
	\includegraphics[width=1.8\columnwidth]{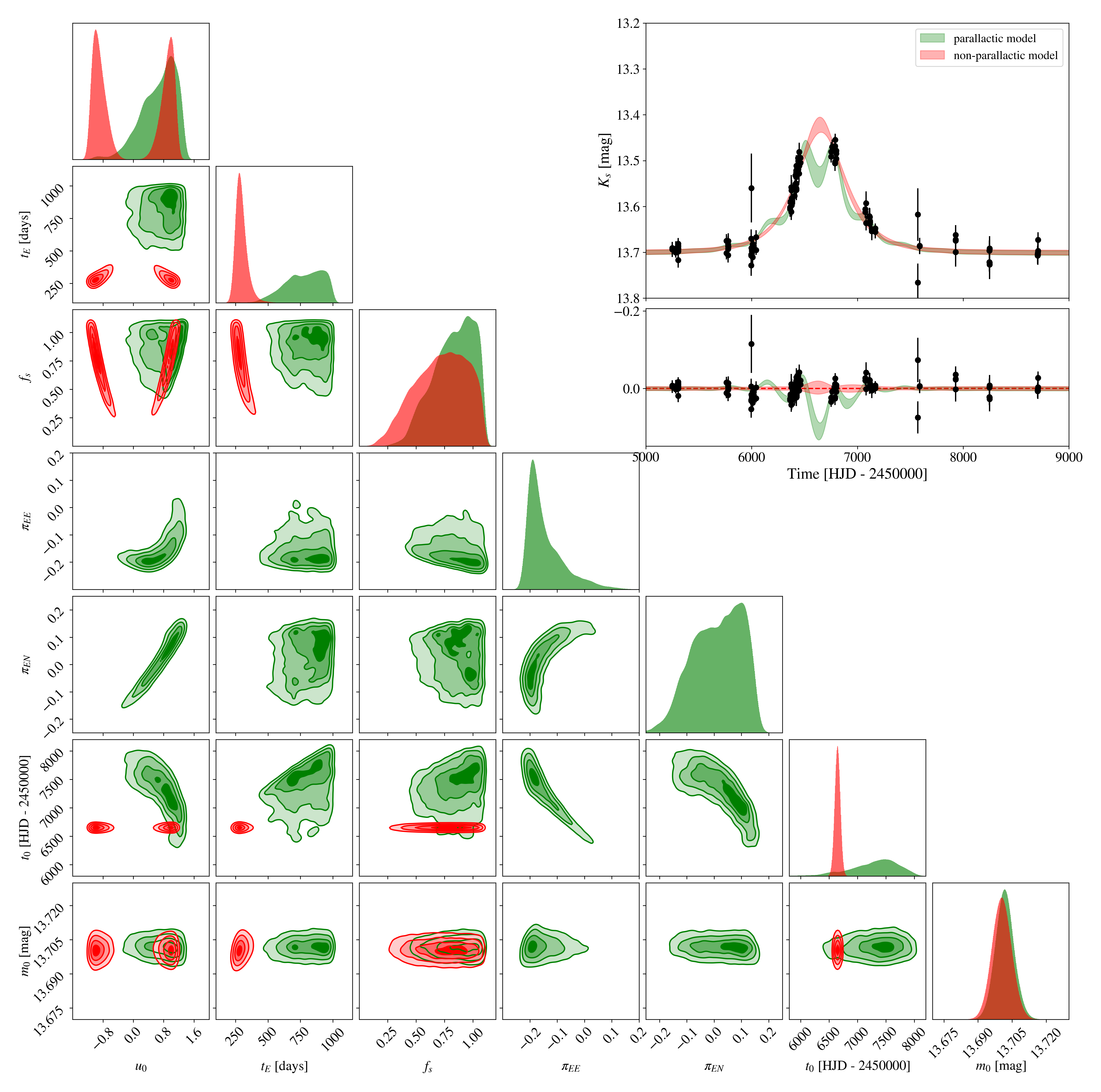}
    \caption{
    As Fig.~\ref{fig:event1_corner_LC}, but for event VVV-2013-DSC-0541.}
        \label{fig:event3_corner_LC}
\end{figure*}
\begin{figure*}
	\includegraphics[width=1.8\columnwidth]{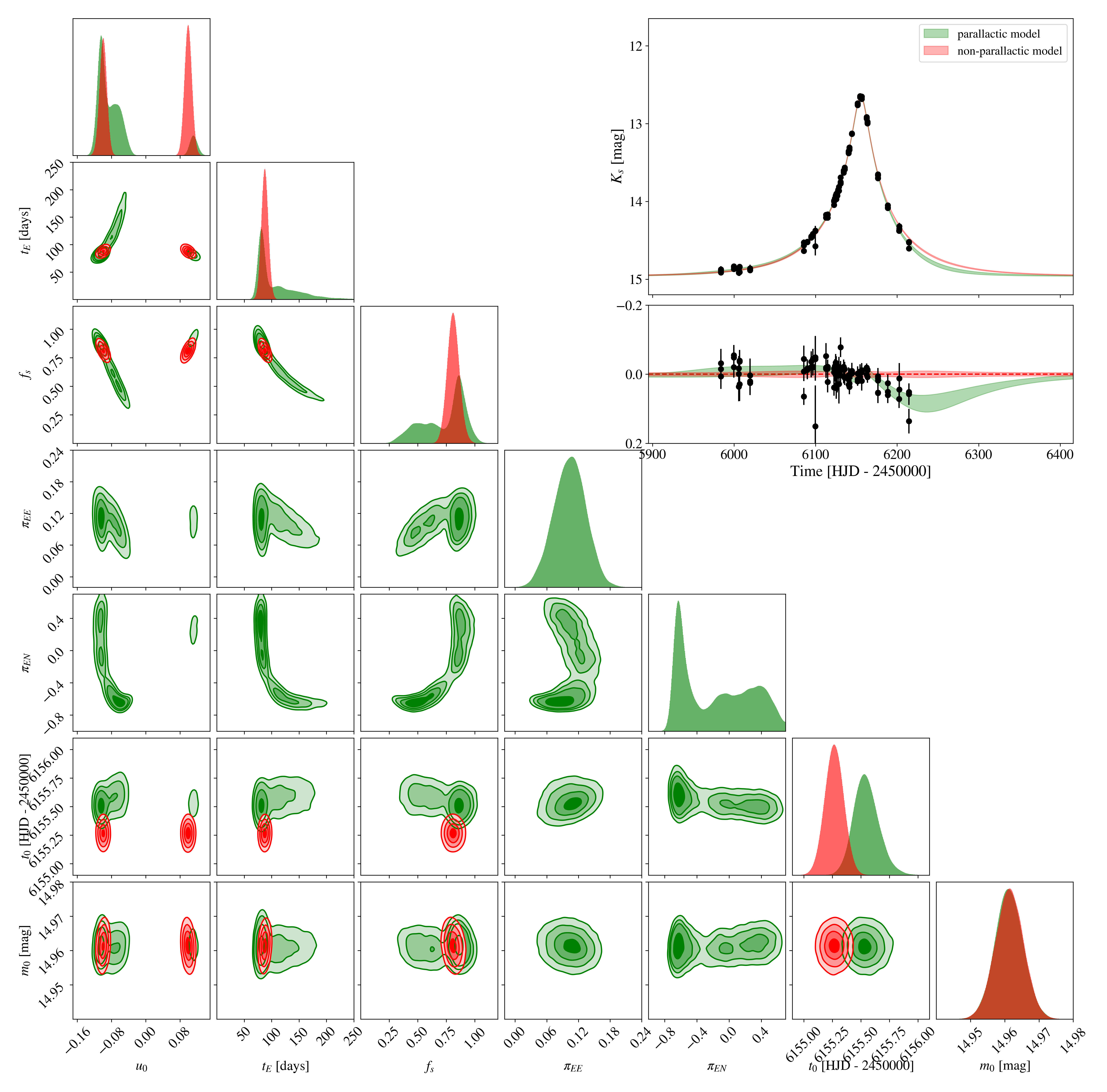}
    \caption{As Fig.~\ref{fig:event1_corner_LC}, but for event VVV-2012-BLG-0440.}
    \label{fig:event4_corner_LC}
\end{figure*}

We now describe in detail the 4 remaining events. We discuss their most probable nature, show the limitations of the analysis that has been done so far and propose follow-up observations where needed. 

\subsubsection{VVV-2013-BLG-0324}

Fig.~\ref{fig:event1_corner_LC} presents the lightcurves for the parallax and non-parallax models for this event, together with the corner plots for the fitted parameters. The blending parameter in the parallax model is estimated to be $f_{\text{s}} = 0.98^{+0.08}_{-0.11}$. So, there is no significant contribution to the observed light by the lens.

The results of recovering the lens mass and distance distribution are presented in Fig.~\ref{fig:mass_distance}. The proper motions used for this analysis were $\mu_{\alpha} = 12.34 \pm 2.76 \textnormal{ mas/yr}$, $\mu_{\delta} = -0.46 \pm 2.47 \textnormal{ mas/yr}$. The lens is a very nearby ($D_{\rm L} = 0.78^{+0.51}_{-0.35}$ kpc) dark remnant -- most probably a neutron star ($M_{\rm L} = 1.46^{+1.13}_{-0.71} \ M_{\odot}$), though possibly a high-mass white dwarf or mass-gap object. The upper and lower estimates of dark lens probability are 0.995 and 0.991 respectively. This analysis was performed using the entire astrometric dataset. However, this source was one of the 3 whose proper motions changed after excluding the amplified points, as discussed in Section~\ref{VIRAC2astrometry}. There remains the possibility of astrometric lensing affecting our proper motion measurements, though this can be ruled out by astrometric follow-up of the source.

With high-precision measurements of the position of the source long after the event took place, we would be able to confirm our values for the source. The alternative is also exciting. If there was a significant difference, we would have detected astrometric lensing through the proper motion anomaly. Knowing the true proper motion of the source, we could model the astrometric deviation in VVV data, as well as the photometry. Simultaneous astrometric and photometric modelling would enable us to measure $\theta_{\rm E}$ and solve the event completely, including the proper motion and mass of the lens.

\subsubsection{VVV-2013-BLG-0460}

In Fig.~\ref{fig:event2_corner_LC}, we present the results of the parallax and non-parallax models for this event. The blending parameter in the parallax model is estimated to be $f_{\text{s}} = 0.91^{+0.12}_{-0.21}$, indicating no significant contribution to the observed light by the lens.

The results of recovering the lens mass and distance distribution are shown in Fig.~\ref{fig:mass_distance}. The proper motions used for input to {\tt Dark Lens} were $\mu_{\alpha} = -2.69 \pm 0.60 \textnormal{ mas/yr}$, $\mu_{\delta} = -0.31 \pm 0.58 \textnormal{ mas/yr}$.
In this event, the lens is located at $D_{\rm L} = 5.26^{+1.46}_{-1.36}$ kpc. It is a dark remnant with a mass of $M_{\rm L} = 1.63^{+1.15}_{-0.70} \ M_{\odot}$ -- similarly to VVV-2013-BLG-0324, it is most likely to be a neutron star (though possibly a high-mass white dwarf or mass-gap object). The upper and lower estimates of dark lens probability are  0.912 and 0.857 respectively.  However, the lens mass-distance distribution in Fig.~\ref{fig:mass_distance} is wide and includes the possibility of the lens residing either in the disc or the bulge. A possible worry with this object is its proper motion is relatively uncertain.
Astrometric follow-up of this object would also be desirable to limit the solution to a smaller volume of the parameter space.

\subsubsection{VVV-2013-DSC-0541}

Although this event made it to the final sample based on its lightcurve and the dark lens probability, its true nature is not clearcut. The cornerplot for the parallax solution in Fig.~\ref{fig:event3_corner_LC} is untypical for a bona fide microlensing event, with an unusually long, loosely constrained timescale $\tE$. Distributions of $u_0$, $t_0$ are also exceptionally wide. The gaps between observation epochs are filled with multiple rises and falls; however, the lightcurve could also potentially be fitted with a very steep rise and slower fall characteristic of some types of intrinsic variables.

Alternative explanations include a nova or a Be star. An argument in favour of the nova hypothesis is the near-IR colours of the source, which are typical for novae \citep[$j - h = 0.466$, $h - k = 0.242$;][]{Saito2013}. However, the lightcurve of the object shows a slow rise over a long timescale ($\Delta K_{\rm s} = 0.2$ mag over $\sim$ 100 days), compared to typical nova outbursts having timescales of hours/days for a 4-15 mag rise ~\citep{Str10}. The incompleteness of data, especially during the putative outburst peak and early decline, prevents us from making a more detailed analysis. Since the amplitude covered by the data is low, it is hard to discuss colour changes. ASAS-SN \citep[All-Sky Automated Survey for Supernovae,][]{Ko17} photometry of this source after the return to baseline (in the assumed microlensing scenario) shows further short-period variability at a level of $\sim$1 mag in the time window between 57500-59500 MJD.
The lightcurves of Be stars can exhibit variability across a range of timescales from hours to decades. Periodicity on shorter timescales of hours to days is typically attributed to stellar pulsations, whilst outbursts and quasi-periodic oscillations are found on intermediate timescales of days to months, though durations of years do sometimes occur~\citep{LB17}. The presence of emission lines in the spectrum is the defining characteristic of this class of object class~\citep{Zo97}. They are believed to originate when either a rapidly rotating B star forms a decretion disc or when there is ongoing mass transfer from a companion through an accretion disc~\citep[see e.g.,][for more discussion]{Be}. Spectroscopic follow-up looking for emission lines can clarify the classification of this source. 

Given its inconclusive nature, we do not reject the possibility of this event being microlensing. The blending parameter in the parallax model is estimated to be $f_{\text{s}} = 0.85^{+0.17}_{-0.21}$, indicating no significant contribution to the observed light by the lens. The proper motions used for the analysis were $\mu_{\alpha} = -6.93 \pm 0.37 \textnormal{ mas/yr}$, $\mu_{\delta} = -0.60 \pm 0.41 \textnormal{ mas/yr}$. If the microlensing hypothesis is true, it is a relatively nearby, ($D_{\rm L} = 2.80^{+2.05}_{-1.52}$ kpc), rather high-mass ($M_{\rm L} = 2.07^{+3.60}_{-1.27} \ M_{\odot}$) remnant. The most likely mass corresponds to a neutron star or a mass-gap object, however the errorbars are large and include other possibilities as well. The upper and lower estimates of dark lens probability are 0.751 and 0.689 respectively.

\subsubsection{VVV-2012-BLG-0440}

This event was originally discovered by \citet{Navarro2017}. Fig.~\ref{fig:event4_corner_LC} shows the results of the parallax and non-parallax models for this event. The blending parameter in the parallax model is $f_{\text{s}} = 0.73^{+0.20}_{-0.30}$. The distribution of $f_{\text{s}}$ is clearly bimodal, with the preferred solution indicating no significant contribution to the observed light by the lens, though the subsidiary mode has the lens contributing the majority of observed light.

The results of recovering the lens mass and distance distribution are presented in Fig.~\ref{fig:mass_distance}. The proper motions used for this analysis were $\mu_{\alpha} = -3.44 \pm 0.48 \textnormal{ mas/yr}$, $\mu_{\delta} = -4.85 \pm 0.50 \textnormal{ mas/yr}$. The recovered values are $D_{\rm L} = 3.70^{+3.67}_{-1.19}$ kpc and $M_{\rm L} = 0.73^{+0.52}_{-0.39} \ M_{\odot}$, with the upper and lower estimates of dark lens probability being 0.555 and 0.216 respectively. This makes the lens a white dwarf candidate, though it could also possibly be a dim main-sequence star.

There are two distinct solutions: one centred on a higher-mass remnant in the disk and the other on a lower-mass one in the bulge. After separating them with a cut at $D_L = 6.5$, we obtain the resulting lens mass and distance: $D_{\rm L, disk} = 3.38^{+1.10}_{-1.01}$ kpc and $M_{\rm L, disk} = 0.83^{+0.49}_{-0.35} \ M_{\odot}$ for the disk solution; $D_{\rm L, bulge} = 7.77^{+0.27}_{-0.41}$ kpc and $M_{\rm L, bulge} = 0.31^{+0.25}_{-0.15} \ M_{\odot}$ for the bulge solution. The disk solution is strongly preferred, as the disk : bulge ratio of the total weight of samples in each subset is 4.3 : 1.

As the source position is very close to the Galactic Centre, the extinction at 8 kpc is high. Due to many samples situated near the dark remnant -- main sequence border, and the uncertainty in drawing this boundary caused by a wide range of possible extinction values, there is a large difference between the upper and lower estimates of dark lens probability. As a way of dealing with this, we also plot two additional lines in Fig.~\ref{fig:mass_distance} as rough estimates of the boundary for extinction at the median distances for the disk ($D_{\rm L \ med, disk} = 3.38$) and bulge ($D_{\rm L \ med, bulge} = 7.77$) solutions, obtained by linear approximation. 

The highly blended solution cannot lead to reliable results. Specifically, if blending is very high, this is contradictory with the assumption that the proper motion fit obtained in Section~\ref{VIRAC2astrometry} is a good estimate of the true proper motion of the source. Redoing the analysis including only the low-blending samples with a cut of $f_{\text{s}} > 0.75$, we find that the bimodality of solutions shown in Fig.~\ref{fig:mass_distance} is not a consequence of blending. The removed high-blending samples have relatively very low weights and were not significantly impacting the solution. Restricting the input to only low-blending samples returns the same mass and distance distributions ($D_{\rm L} = 3.71^{+3.69}_{-1.19}$ kpc and $M_{\rm L} = 0.72^{+0.52}_{-0.38} \ M_{\odot}$). The lens mass -- distance and blend light -- lens light plots also remain visually unchanged, so we do not show them for low-blending samples separately.

With the data available, we are not able to distinguish between the low blending case (that leads to the bimodal disk/bulge solution described above in detail), or the high blending case (in which we cannot study the event more closely, because we do not have reliable estimates of the source proper motion). The solution to this problem is to conduct high resolution follow-up observations aiming to either resolve the source and the luminous lens in the high blending case or confirm the low blending assumption with a non-detection of light from the lens at an expected threshold. Both of those results would be scientifically interesting, as one would give more evidence for the lens being a dark stellar remnant, and in the other, the lens-source relative proper motion $\mu_{\rm rel}$ could be measured and the event completely solved.

\section{Conclusions}

We carried out a search for dark lenses in the VISTA Variables in the Via Lactea (VVV) survey. Our tool of choice is parallax microlensing, in which the annual motion of the Earth is detectable via a distortion of the standard lightcurve. Measurement of the microlensing parallax imposes additional constraints on the mass of the lens, beyond standard microlensing.

We report the results of using the nested sampling method to automatically fit simple and parallax microlensing models to a sample of 1959 candidate events. These were previously identified by \citet{Husseiniova2021} in a systematic search through 700 million VVV lightcurves. Given the low cadence and noise properties of the VVV, the use of nested sampling is not just desirable but essential to probe the degeneracies in the model fits. We use likelihoods of those fits to automatically identify candidates for parallax lensing events. By applying this method, we greatly reduce the time needed for conducting such an analysis. We apply reliable statistical measures for candidate selection, eliminate the need for human supervision, and automatically recover multimodal solutions and their relative likelihoods. This emphasises the advantage of our approach compared to the applications of Markov Chain Monte Carlo (MCMC) methods routinely used in previous works. We make the codes and data used in this search publicly available to use in future studies.

This yields 176 events for which the Bayes factor strongly prefers the microlensing parallax model over standard microlensing. We then select a smaller subsample of 21 of the most promising events for further analysis. Here, a very modest amount of human intervention is required, though this is ameliorated by use of multiple assessors to judge the candidates. For these 21 events, we extract the position and proper motions of the source by refitting the VVV time series of positional data. This necessitates care to minimise contamination from the effects of astrometric microlensing itself. (Simultaneous microlensing fits to both the VVV astrometry and photometry were explored but the data quality were insufficient to yield definitive results.)  

With the source proper motions in hand, we use a model of the Galaxy with deflector populations in both Bulge and disc to derive the probability of the mass and distance of the lens given the microlensing model data. By computing the brightness of a main sequence star at the lens distance and comparing to the blending parameters of the photometric fit, we can estimate the probability of a dark lens~\citep{Wyrzykowski2016,Mroz2021a}. This gives us probability density plots for the lens mass and distance as well as the blend light and lens light.

In the end, we obtain 4 candidates for parallax microlensing events caused by dark stellar remnants. The best candidate is a nearby ($D_{\rm L} = 0.78^{+0.51}_{-0.35}$ kpc) stellar remnant with a mass of $M_{\rm L} = 1.46^{+1.13}_{-0.71} \ M_{\odot}$. The second best is located at $D_{\rm L} = 5.26^{+1.46}_{-1.36}$ kpc and has a mass of $M_{\rm L} = 1.63^{+1.15}_{-0.70} \ M_{\odot}$. Both of those candidates are most probably neutron stars, though high mass white dwarfs remain still possible. For the remaining 2 candidates, limitations in the data prevent us from being more categorical in our assessments. One may be a relatively nearby ($D_{\rm L} = 2.80^{+2.05}_{-1.52}$ kpc), high-mass ($M_{\rm L} = 2.07^{+3.60}_{-1.27} \ M_{\odot}$) remnant. However, the blue colour and persistent baseline variability of the source -- as well as the inferences of microlensing parameters including the very wide distributions and an unusually long timescale -- suggest a Be star is also a viable possibility. The other is confidently a parallax microlensing event, but the lens may be dark or luminous. This is because the extinction at the source location close to the Galactic Centre is both high and uncertain.

Follow-up observations can confirm the dark lens nature of our best candidates and recover their mass and distance more accurately. For example, NIR follow-up of these events with the {\it Hubble Space Telescope} (HST) can pin down the source of the blending, and perhaps even resolve the lens and source if they have had enough time to separate. The VVV data have low cadence relative to surveys dedicated to microlensing. This has limited our analysis and hence the certainty of our candidates. Future observatories and missions observing in the near infrared, in particular {\it The Vera C. Rubin Observatory} and {\it The Roman Space Telescope}, will provide much more precise photometry and astrometry and allow for identifying many more candidates in those regions of the Galaxy.

To give an idea of the possibilities, we discuss a hypothetical event resembling our most promising candidate, as seen by {\it Roman}. We create mock photometric and astrometric data for this object, choosing the more preferred of the degenerate solutions ($u_0>0$). We take the median values of each parameter from the photometric parameter set ($t_{\text{\rm E}}, u_{0}, t_{0}, m_{0}, f_{\text{s}}, \pi_{\text{EE}}, \pi_{\text{EN}}$) for all $u_0>0$ samples as input values. We then calculate the remaining parameters needed to simulate astrometry. We use the median lens mass from all $u_0>0$ samples (1.4773 $M_{\odot}$) in the output of the {\tt Dark Lens} code to calculate the input value of $\theta_{\rm E}$ (from Equation \ref{tE}; $\theta_{\rm E} = \kappa M \pi_{\rm E}$). We also use the proper motions of the source obtained from the fits to VIRAC2 time-series data. We assume that the parallax deviation is to be attributed fully to the motion of the lens, and simulate a straight-line motion for the source. Finally, we take the reference position -- the position the source would be in at $t=t_{0}$, if there was no lensing -- to be (0,0).

With this set of parameters, we simulate the photometry and astrometry of this event, and apply the typical {\it Roman} cadence \citep[][]{Gaudi2019}\footnote{\url{https://roman.gsfc.nasa.gov/galactic_bulge_time_domain_survey.html}}.
{\it Roman} will observe fields in the Galactic Bulge in 72-day seasons; for the entire duration of each season, the cadence will be 15 minutes. There will be six seasons in total - three seasons, each separated by 1/2 year, at the beginning and similarly, three at the end of the mission, which will have a nominal length of 5 years. To fix the timescales so that the three first seasons happen during amplification, we take MJD = 56300 as a starting date for this observing pattern.

The precision of those observations as a function of magnitude is not yet known, and will be analysed during the operations of the mission. We take the expected astrometric accuracy as 1 mas and photometric accuracy as 10 milli-mag~\citep[following][]{Penny2019, WFIRST2019, Gaudi2019}. These are values appropriate for $\approx$ 21 mag stars (in W149/$H_{AB}$ bands), which is very dim in comparison to our source ($K_s = 16.8$ at baseline). This implies {\it Roman} will be able to detect and study far more events than available to us now. We add Gaussian noise to astrometric position and magnitude at each epoch to create our mock dataset.

\begin{figure*}
	\includegraphics[width=1.8\columnwidth]{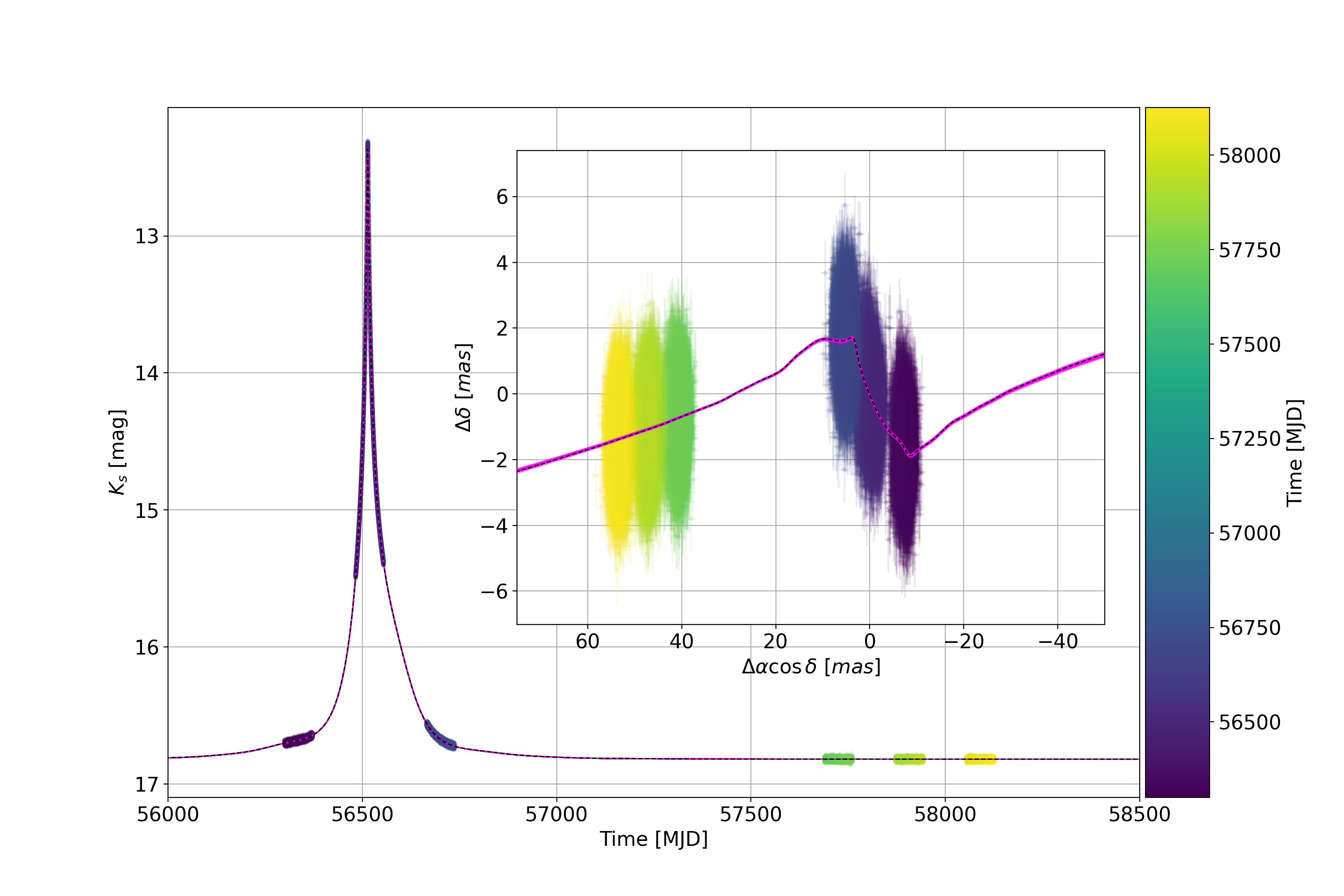}
    \caption{Photometry and astrometry of the event VVV-2013-BLG-0324 as it would be seen by the future {\it Roman Space Telescope} mission (assuming one of our solutions). The angular Einstein radius $\thetaE$ of the event in this simulation is 6.4 mas. Main figure: simulated lightcurve of the event. Top right: simulated astrometric track of the light center in the event. Dashed black lines represent the simulated true lightcurve and astrometric track, low-alpha dots with errorbars represent the mock {\it Roman} datapoints, and solid magenta lines represent 100 randomly chosen samples from the posterior distributions.}
    \label{fig:rst_astrom_photom}
\end{figure*}

\begin{figure*}
	\includegraphics[width=1.8\columnwidth]{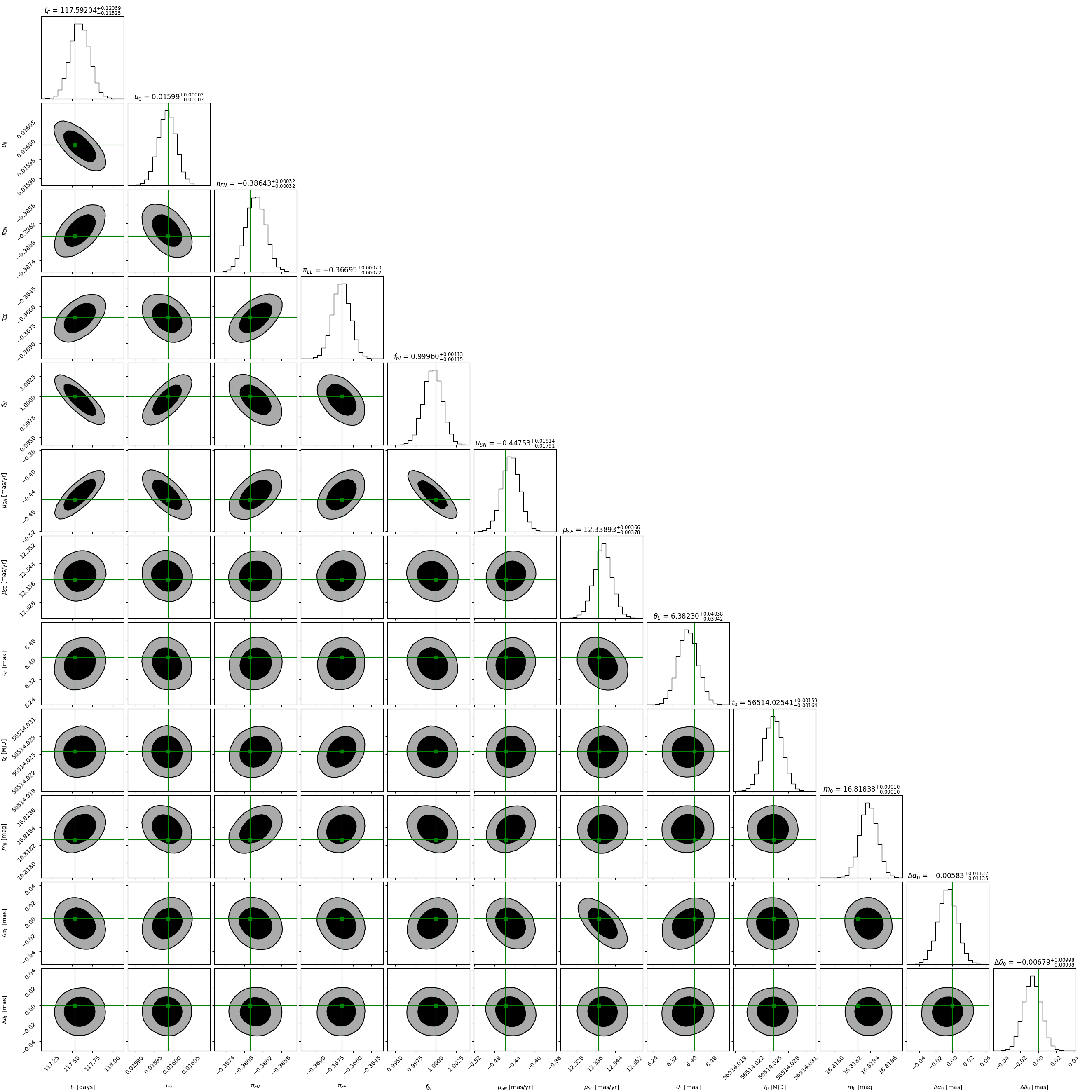}
    \caption{Corner plot of posterior distributions of photometric and astrometric lensing parameters resulting from modelling the mock data shown in Figure \ref{fig:rst_astrom_photom}. Contours indicate 1$\sigma$ and 2$\sigma$ levels. Green lines indicate true input values used for the simulation.}
    \label{fig:rst_fits}
\end{figure*}

We then perform a simultaneous fit to photometric and astrometric data, similarly to \citet{Rybicki2018}. Unlike \citet{Rybicki2018}, we also fit for the unknown reference position ($\Delta\alpha_{0}, \Delta\delta_{0}$). In our parametrisation, we use the geocentric parameter set ($t_{\text{\rm E}}, u_{0}, t_{0}, m_{0}, f_{\text{s}}, \pi_{\text{EE}}, \pi_{\text{EN}}$), with additional astrometric parameters ($\mu_{\text{SE}}, \mu_{\text{SN}}, \theta_{\text{E}}, \Delta\alpha_{0}, \Delta\delta_{0}$), for consistency with our methods described in 3.1. The nested sampling settings and priors were the same as in Table \ref{table:priors}, with the addition of priors for the astrometric parameters. We do not include separate fitting for astrometric deviations from straight-line motion of the source, which would introduce additional shifts at a $\leq 0.125$ mas level to the datapoints for a typical source situated in the Galactic Bulge. We present the results of this experiment in the following two figures: we show the simulated and fitted lightcurve and astrometric track in Fig.~\ref{fig:rst_astrom_photom}, and the posterior distributions in Fig.~\ref{fig:rst_fits}.

\begin{table}
\centering
\begin{tabular}{lll} 
\hline
parameter & prior & unit \\
 \hline
 $\mu_{\text{SE}}$ & uniform(-15,15) & mas/yr \\
 $\mu_{\text{SN}}$ & uniform(-15,15) & mas/yr \\
 $\theta_{\text{E}}$ & uniform(0,10) & mas \\
 $\Delta\alpha_{0}$ & uniform($\Delta\alpha_{\text{min}}$, $\Delta\alpha_{\text{max}}$) & mas \\
 $\Delta\delta_{0}$ & uniform($\Delta\delta_{\text{min}}$, $\Delta\delta_{\text{max}}$) & mas \\
 
\end{tabular}
\caption{Priors for additional astrometric parameters used in the modelling of the mock data. $\Delta\alpha_{\text{min}},\Delta\delta_{\text{min}}$ and $\Delta\alpha_{\text{max}},\Delta\delta_{\text{max}}$ are the minimum and maximum values found in the dataset.}
\label{table:priors2}
\end{table}

We are able to accurately recover the true solution, breaking the degeneracy that is found in photometric microlensing alone. From the recovered distributions of $\theta_E = 6.382^{+0.040}_{-0.039}$, and $\pi_E = 0.53291 ^{+0.0066}_{-0.0068}$, we directly calculate the lens mass to be $M_{lens} = 1.4715 ^{+0.0095}_{-0.0096} M_{\odot}$, which is consistent with the input value of 1.4773 $M_{\odot}$ (all reported values are median values from all resulting samples, with 84th-50th percentile indicated as a superscript and 50th-16th percentile indicated as a subscript).

The lens mass measurement, unlike the mass inference using the {\tt Dark Lens} code shown in this work, can now be done without any assumptions about the distance and velocity distribution for the lens, and from the astrometric and photometric data alone. We conclude that possibilities that will become available in the next decade will allow for studying a much larger sample of events, and conducting a much more advanced analysis. With simultaneous measurements of $\boldsymbol{\pi}_{\rm E}$
and $\theta_{\rm E}$, it will be possible to recover lens masses and proper motions in a straightforward way, and with a precision by far exceeding that of methods available today.

\section*{Acknowledgements}
We would like to thank Roberto Saito for help with understanding the nature of VVV-2013-DSC-0541. We would like to thank K. A. Rybicki for helpful discussions on the future {\it Roman} mission. We received a helpful report from the referee (Nicholas Rattenbury).

\section*{Data Availability}
The electronic table of parallax microlensing events with associated photometry and astrometry is published with this article as online supplementary material. The codes used for modelling photometry and astrometry of microlensing events are publicly available at \url{https://github.com/zofiakaczmarek/nested_ulens_parallax}.
%%%%%%%%%%%%%%%%%%%%%%%%%%%%%%%%%%%%%%%%%%%%%%%%%%


\bibliographystyle{mnras}
\bibliography{vvv}

\begin{thebibliography}{}
\makeatletter
\relax
\def\mn@urlcharsother{\let\do\@makeother \do\$\do\&\do\#\do\^\do\_\do\%\do\~}
\def\mn@doi{\begingroup\mn@urlcharsother \@ifnextchar [ {\mn@doi@}
  {\mn@doi@[]}}
\def\mn@doi@[#1]#2{\def\@tempa{#1}\ifx\@tempa\@empty \href
  {http://dx.doi.org/#2} {doi:#2}\else \href {http://dx.doi.org/#2} {#1}\fi
  \endgroup}
\def\mn@eprint#1#2{\mn@eprint@#1:#2::\@nil}
\def\mn@eprint@arXiv#1{\href {http://arxiv.org/abs/#1} {{\tt arXiv:#1}}}
\def\mn@eprint@dblp#1{\href {http://dblp.uni-trier.de/rec/bibtex/#1.xml}
  {dblp:#1}}
\def\mn@eprint@#1:#2:#3:#4\@nil{\def\@tempa {#1}\def\@tempb {#2}\def\@tempc
  {#3}\ifx \@tempc \@empty \let \@tempc \@tempb \let \@tempb \@tempa \fi \ifx
  \@tempb \@empty \def\@tempb {arXiv}\fi \@ifundefined
  {mn@eprint@\@tempb}{\@tempb:\@tempc}{\expandafter \expandafter \csname
  mn@eprint@\@tempb\endcsname \expandafter{\@tempc}}}

\bibitem[\protect\citeauthoryear{{Abbott} et~al.,}{{Abbott}
  et~al.}{2020}]{Abbott2020}
{Abbott} R.,  et~al., 2020, arXiv e-prints, \href
  {https://ui.adsabs.harvard.edu/abs/2020arXiv200408342T} {p. arXiv:2004.08342}

\bibitem[\protect\citeauthoryear{{Alcock} et~al.,}{{Alcock}
  et~al.}{1995}]{Alcock1995}
{Alcock} C.,  et~al., 1995, \mn@doi [\apjl] {10.1086/309783}, \href
  {https://ui.adsabs.harvard.edu/abs/1995ApJ...454L.125A} {454, L125}

\bibitem[\protect\citeauthoryear{{Alonso-Garc{\'\i}a}, {Mateo}, {Sen},
  {Banerjee}, {Catelan}, {Minniti}  \& {von Braun}}{{Alonso-Garc{\'\i}a}
  et~al.}{2012}]{dophot2}
{Alonso-Garc{\'\i}a} J.,  {Mateo} M.,  {Sen} B.,  {Banerjee} M.,  {Catelan} M.,
   {Minniti} D.,   {von Braun} K.,  2012, \mn@doi [\aj]
  {10.1088/0004-6256/143/3/70}, \href
  {https://ui.adsabs.harvard.edu/abs/2012AJ....143...70A} {143, 70}

\bibitem[\protect\citeauthoryear{{Astropy Collaboration} et~al.,}{{Astropy
  Collaboration} et~al.}{2013}]{Astropy2013}
{Astropy Collaboration} et~al., 2013, \mn@doi [\aap]
  {10.1051/0004-6361/201322068}, \href
  {https://ui.adsabs.harvard.edu/abs/2013A&A...558A..33A} {558, A33}

\bibitem[\protect\citeauthoryear{{Astropy Collaboration} et~al.,}{{Astropy
  Collaboration} et~al.}{2018}]{Astropy2018}
{Astropy Collaboration} et~al., 2018, \mn@doi [\aj] {10.3847/1538-3881/aabc4f},
  \href {https://ui.adsabs.harvard.edu/abs/2018AJ....156..123A} {156, 123}

\bibitem[\protect\citeauthoryear{{Belokurov} \& {Evans}}{{Belokurov} \&
  {Evans}}{2002}]{Belokurov2002}
{Belokurov} V.~A.,  {Evans} N.~W.,  2002, \mn@doi [\mnras]
  {10.1046/j.1365-8711.2002.05222.x}, \href
  {https://ui.adsabs.harvard.edu/abs/2002MNRAS.331..649B} {331, 649}

\bibitem[\protect\citeauthoryear{{Bennett} et~al.,}{{Bennett}
  et~al.}{2002}]{Be02}
{Bennett} D.~P.,  et~al., 2002, \mn@doi [\apj] {10.1086/342225}, \href
  {https://ui.adsabs.harvard.edu/abs/2002ApJ...579..639B} {579, 639}

\bibitem[\protect\citeauthoryear{{Boubert} \& {Evans}}{{Boubert} \&
  {Evans}}{2018}]{Be}
{Boubert} D.,  {Evans} N.~W.,  2018, \mn@doi [\mnras] {10.1093/mnras/sty980},
  \href {https://ui.adsabs.harvard.edu/abs/2018MNRAS.477.5261B} {477, 5261}

\bibitem[\protect\citeauthoryear{{Carr}, {K{\"u}hnel}  \& {Sandstad}}{{Carr}
  et~al.}{2016}]{Ca16}
{Carr} B.,  {K{\"u}hnel} F.,   {Sandstad} M.,  2016, \mn@doi [\prd]
  {10.1103/PhysRevD.94.083504}, \href
  {https://ui.adsabs.harvard.edu/abs/2016PhRvD..94h3504C} {94, 083504}

\bibitem[\protect\citeauthoryear{{Cassan} et~al.,}{{Cassan}
  et~al.}{2021}]{Cassan2021}
{Cassan} A.,  et~al., 2021, \mn@doi [Nature Astronomy]
  {10.1038/s41550-021-01514-w}, \href
  {https://ui.adsabs.harvard.edu/abs/2021NatAs.tmp..230C} {}

\bibitem[\protect\citeauthoryear{{Chwolson}}{{Chwolson}}{1924}]{Chwolson1924}
{Chwolson} O.,  1924, \mn@doi [Astronomische Nachrichten]
  {10.1002/asna.19242212003}, \href
  {https://ui.adsabs.harvard.edu/abs/1924AN....221..329C} {221, 329}

\bibitem[\protect\citeauthoryear{{Dominik} \& {Sahu}}{{Dominik} \&
  {Sahu}}{2000}]{DoSa}
{Dominik} M.,  {Sahu} K.~C.,  2000, \mn@doi [\apj] {10.1086/308716}, \href
  {https://ui.adsabs.harvard.edu/abs/2000ApJ...534..213D} {534, 213}

\bibitem[\protect\citeauthoryear{{Dong} et~al.,}{{Dong}
  et~al.}{2019}]{Dong2019}
{Dong} S.,  et~al., 2019, \mn@doi [\apj] {10.3847/1538-4357/aaeffb}, \href
  {https://ui.adsabs.harvard.edu/abs/2019ApJ...871...70D} {871, 70}

\bibitem[\protect\citeauthoryear{{Einstein}}{{Einstein}}{1936}]{Einstein1936}
{Einstein} A.,  1936, \mn@doi [Science] {10.1126/science.84.2188.506}, \href
  {https://ui.adsabs.harvard.edu/abs/1936Sci....84..506E} {84, 506}

\bibitem[\protect\citeauthoryear{{Evans}}{{Evans}}{2003}]{Evans2003}
{Evans} N.~W.,  2003, arXiv e-prints, \href
  {https://ui.adsabs.harvard.edu/abs/2003astro.ph..4252E} {pp
  astro--ph/0304252}

\bibitem[\protect\citeauthoryear{{Evans} \& {Belokurov}}{{Evans} \&
  {Belokurov}}{2002}]{EB}
{Evans} N.~W.,  {Belokurov} V.,  2002, \mn@doi [\apjl] {10.1086/340004}, \href
  {https://ui.adsabs.harvard.edu/abs/2002ApJ...567L.119E} {567, L119}

\bibitem[\protect\citeauthoryear{{Foreman-Mackey}, {Hogg}, {Lang}  \&
  {Goodman}}{{Foreman-Mackey} et~al.}{2013}]{Foreman-Mackey2013}
{Foreman-Mackey} D.,  {Hogg} D.~W.,  {Lang} D.,   {Goodman} J.,  2013, \mn@doi
  [\pasp] {10.1086/670067}, \href
  {https://ui.adsabs.harvard.edu/abs/2013PASP..125..306F} {125, 306}

\bibitem[\protect\citeauthoryear{{Gaia Collaboration} et~al.,}{{Gaia
  Collaboration} et~al.}{2018}]{GDR22018}
{Gaia Collaboration} et~al., 2018, \mn@doi [\aap]
  {10.1051/0004-6361/201833051}, \href
  {https://ui.adsabs.harvard.edu/abs/2018A&A...616A...1G} {616, A1}

\bibitem[\protect\citeauthoryear{{Gaia Collaboration} et~al.,}{{Gaia
  Collaboration} et~al.}{2021}]{GEDR32021}
{Gaia Collaboration} et~al., 2021, \mn@doi [\aap]
  {10.1051/0004-6361/202039657}, \href
  {https://ui.adsabs.harvard.edu/abs/2021A&A...649A...1G} {649, A1}

\bibitem[\protect\citeauthoryear{{Garc{\'\i}a-Bellido} \&
  {Clesse}}{{Garc{\'\i}a-Bellido} \& {Clesse}}{2018}]{GB2018b}
{Garc{\'\i}a-Bellido} J.,  {Clesse} S.,  2018, \mn@doi [Physics of the Dark
  Universe] {10.1016/j.dark.2018.01.001}, \href
  {https://ui.adsabs.harvard.edu/abs/2018PDU....19..144G} {19, 144}

\bibitem[\protect\citeauthoryear{{Garc{\'\i}a-Bellido}, {Clesse}  \&
  {Fleury}}{{Garc{\'\i}a-Bellido} et~al.}{2018}]{GB2018a}
{Garc{\'\i}a-Bellido} J.,  {Clesse} S.,   {Fleury} P.,  2018, \mn@doi [Physics
  of the Dark Universe] {10.1016/j.dark.2018.04.005}, \href
  {https://ui.adsabs.harvard.edu/abs/2018PDU....20...95G} {20, 95}

\bibitem[\protect\citeauthoryear{{Gaudi} et~al.,}{{Gaudi}
  et~al.}{2019}]{Gaudi2019}
{Gaudi} B.~S.,  et~al., 2019, \baas, \href
  {https://ui.adsabs.harvard.edu/abs/2019BAAS...51c.211G} {51, 211}

\bibitem[\protect\citeauthoryear{{Golovich} et~al.,}{{Golovich}
  et~al.}{2020}]{Golovich2020}
{Golovich} N.,  et~al., 2020, arXiv e-prints, \href
  {https://ui.adsabs.harvard.edu/abs/2020arXiv200907927G} {p. arXiv:2009.07927}

\bibitem[\protect\citeauthoryear{{Gould}}{{Gould}}{1995}]{Gould1995}
{Gould} A.,  1995, \mn@doi [\apjl] {10.1086/187933}, \href
  {https://ui.adsabs.harvard.edu/abs/1995ApJ...446L..71G} {446, L71}

\bibitem[\protect\citeauthoryear{{Gould}}{{Gould}}{2000a}]{Gould2000b}
{Gould} A.,  2000a, \mn@doi [\apj] {10.1086/308865}, \href
  {https://ui.adsabs.harvard.edu/abs/2000ApJ...535..928G} {535, 928}

\bibitem[\protect\citeauthoryear{{Gould}}{{Gould}}{2000b}]{Gould2000}
{Gould} A.,  2000b, \mn@doi [\apj] {10.1086/317037}, \href
  {https://ui.adsabs.harvard.edu/abs/2000ApJ...542..785G} {542, 785}

\bibitem[\protect\citeauthoryear{{Gould}}{{Gould}}{2004}]{Gould2004}
{Gould} A.,  2004, \mn@doi [\apj] {10.1086/382782}, \href
  {https://ui.adsabs.harvard.edu/abs/2004ApJ...606..319G} {606, 319}

\bibitem[\protect\citeauthoryear{{Green}}{{Green}}{2016}]{Gr16}
{Green} A.~M.,  2016, \mn@doi [\prd] {10.1103/PhysRevD.94.063530}, \href
  {https://ui.adsabs.harvard.edu/abs/2016PhRvD..94f3530G} {94, 063530}

\bibitem[\protect\citeauthoryear{{Griest} et~al.,}{{Griest}
  et~al.}{1991}]{Griest1991}
{Griest} K.,  et~al., 1991, \mn@doi [\apjl] {10.1086/186028}, \href
  {http://adsabs.harvard.edu/abs/1991ApJ...372L..79G} {372, L79}

\bibitem[\protect\citeauthoryear{Higson, Handley, Hobson  \& Lasenby}{Higson
  et~al.}{2019}]{Higson2019}
Higson E.,  Handley W.,  Hobson M.,   Lasenby A.,  2019, Statistics and
  Computing, 29, 891

\bibitem[\protect\citeauthoryear{{Husseiniova}, {McGill}, {Smith}  \&
  {Evans}}{{Husseiniova} et~al.}{2021}]{Husseiniova2021}
{Husseiniova} A.,  {McGill} P.,  {Smith} L.~C.,   {Evans} N.~W.,  2021, \mn@doi
  [\mnras] {10.1093/mnras/stab1882}, \href
  {https://ui.adsabs.harvard.edu/abs/2021MNRAS.506.2482H} {506, 2482}

\bibitem[\protect\citeauthoryear{{Irwin} et~al.,}{{Irwin} et~al.}{2004}]{vdfs}
{Irwin} M.~J.,  et~al., 2004, {VISTA data flow system: pipeline processing for
  WFCAM and VISTA}.
pp 411--422, \mn@doi{10.1117/12.551449}

\bibitem[\protect\citeauthoryear{{Jonker}, {Kaur}, {Stone}  \&
  {Torres}}{{Jonker} et~al.}{2021}]{Jonker2021}
{Jonker} P.~G.,  {Kaur} K.,  {Stone} N.,   {Torres} M. A.~P.,  2021, \mn@doi
  [\apj] {10.3847/1538-4357/ac2839}, \href
  {https://ui.adsabs.harvard.edu/abs/2021ApJ...921..131J} {921, 131}

\bibitem[\protect\citeauthoryear{{Karolinski} \& {Zhu}}{{Karolinski} \&
  {Zhu}}{2020}]{Karolinski2020}
{Karolinski} N.,  {Zhu} W.,  2020, \mn@doi [\mnras] {10.1093/mnrasl/slaa121},
  \href {https://ui.adsabs.harvard.edu/abs/2020MNRAS.498L..25K} {498, L25}

\bibitem[\protect\citeauthoryear{{Kashlinsky}}{{Kashlinsky}}{2016}]{Ka16}
{Kashlinsky} A.,  2016, \mn@doi [\apjl] {10.3847/2041-8205/823/2/L25}, \href
  {https://ui.adsabs.harvard.edu/abs/2016ApJ...823L..25K} {823, L25}

\bibitem[\protect\citeauthoryear{{Kim} et~al.,}{{Kim}
  et~al.}{2016}]{KMTnet2016}
{Kim} S.-L.,  et~al., 2016, \mn@doi [Journal of Korean Astronomical Society]
  {10.5303/JKAS.2016.49.1.037}, \href
  {https://ui.adsabs.harvard.edu/abs/2016JKAS...49...37K} {49, 37}

\bibitem[\protect\citeauthoryear{{Kochanek} et~al.,}{{Kochanek}
  et~al.}{2017}]{Ko17}
{Kochanek} C.~S.,  et~al., 2017, \mn@doi [\pasp] {10.1088/1538-3873/aa80d9},
  \href {https://ui.adsabs.harvard.edu/abs/2017PASP..129j4502K} {129, 104502}

\bibitem[\protect\citeauthoryear{{Koz{\l}owski}, {Wo{\'z}niak}, {Mao}  \&
  {Wood}}{{Koz{\l}owski} et~al.}{2007}]{Kozlowski2007}
{Koz{\l}owski} S.,  {Wo{\'z}niak} P.~R.,  {Mao} S.,   {Wood} A.,  2007, \mn@doi
  [\apj] {10.1086/522563}, \href
  {https://ui.adsabs.harvard.edu/abs/2007ApJ...671..420K} {671, 420}

\bibitem[\protect\citeauthoryear{{Kruszy{\'n}ska} et~al.,}{{Kruszy{\'n}ska}
  et~al.}{2021}]{Kruszynska2021}
{Kruszy{\'n}ska} K.,  et~al., 2021, arXiv e-prints, \href
  {https://ui.adsabs.harvard.edu/abs/2021arXiv211108337K} {p. arXiv:2111.08337}

\bibitem[\protect\citeauthoryear{{Labadie-Bartz} et~al.,}{{Labadie-Bartz}
  et~al.}{2017}]{LB17}
{Labadie-Bartz} J.,  et~al., 2017, \mn@doi [\aj] {10.3847/1538-3881/aa6396},
  \href {https://ui.adsabs.harvard.edu/abs/2017AJ....153..252L} {153, 252}

\bibitem[\protect\citeauthoryear{{Lam}, {Lu}, {Hosek}, {Dawson}  \&
  {Golovich}}{{Lam} et~al.}{2020}]{Lam2020}
{Lam} C.~Y.,  {Lu} J.~R.,  {Hosek} Matthew~W. J.,  {Dawson} W.~A.,   {Golovich}
  N.~R.,  2020, \mn@doi [\apj] {10.3847/1538-4357/ab5fd3}, \href
  {https://ui.adsabs.harvard.edu/abs/2020ApJ...889...31L} {889, 31}

\bibitem[\protect\citeauthoryear{{Lam} et~al.,}{{Lam} et~al.}{2022}]{Lam2022}
{Lam} C.~Y.,  et~al., 2022, arXiv e-prints, \href
  {https://ui.adsabs.harvard.edu/abs/2022arXiv220201903L} {p. arXiv:2202.01903}

\bibitem[\protect\citeauthoryear{{Lee}}{{Lee}}{2017}]{Lee2017}
{Lee} C.-H.,  2017, \mn@doi [Universe] {10.3390/universe3030053}, \href
  {https://ui.adsabs.harvard.edu/abs/2017Univ....3...53L} {3, 53}

\bibitem[\protect\citeauthoryear{{Mao} et~al.,}{{Mao} et~al.}{2002}]{Ma02}
{Mao} S.,  et~al., 2002, \mn@doi [\mnras] {10.1046/j.1365-8711.2002.04986.x},
  \href {https://ui.adsabs.harvard.edu/abs/2002MNRAS.329..349M} {329, 349}

\bibitem[\protect\citeauthoryear{{McGill}, {Smith}, {Evans}, {Belokurov}  \&
  {Smart}}{{McGill} et~al.}{2018}]{McGill2018}
{McGill} P.,  {Smith} L.~C.,  {Evans} N.~W.,  {Belokurov} V.,   {Smart} R.~L.,
  2018, \mn@doi [\mnras] {10.1093/mnrasl/sly066}, \href
  {https://ui.adsabs.harvard.edu/abs/2018MNRAS.478L..29M} {478, L29}

\bibitem[\protect\citeauthoryear{{McGill}, {Smith}, {Evans}, {Belokurov}  \&
  {Zhang}}{{McGill} et~al.}{2019a}]{McGill2019}
{McGill} P.,  {Smith} L.~C.,  {Evans} N.~W.,  {Belokurov} V.,   {Zhang} Z.~H.,
  2019a, \mn@doi [\mnras] {10.1093/mnras/sty3344}, \href
  {https://ui.adsabs.harvard.edu/abs/2019MNRAS.483.4210M} {483, 4210}

\bibitem[\protect\citeauthoryear{{McGill}, {Smith}, {Evans}, {Belokurov}  \&
  {Lucas}}{{McGill} et~al.}{2019b}]{McGill2019b}
{McGill} P.,  {Smith} L.~C.,  {Evans} N.~W.,  {Belokurov} V.,   {Lucas} P.~W.,
  2019b, \mn@doi [\mnras] {10.1093/mnrasl/slz073}, \href
  {https://ui.adsabs.harvard.edu/abs/2019MNRAS.487L...7M} {487, L7}

\bibitem[\protect\citeauthoryear{{Minniti} et~al.,}{{Minniti}
  et~al.}{2010}]{Minniti2010}
{Minniti} D.,  et~al., 2010, \mn@doi [\na] {10.1016/j.newast.2009.12.002},
  \href {http://adsabs.harvard.edu/abs/2010NewA...15..433M} {15, 433}

\bibitem[\protect\citeauthoryear{{Mr{\'o}z} \& {Wyrzykowski}}{{Mr{\'o}z} \&
  {Wyrzykowski}}{2021}]{Mroz2021a}
{Mr{\'o}z} P.,  {Wyrzykowski} {\L}.,  2021, \mn@doi [\actaa]
  {10.32023/0001-5237/71.2.1}, \href
  {https://ui.adsabs.harvard.edu/abs/2021AcA....71...89M} {71, 89}

\bibitem[\protect\citeauthoryear{{Mr{\'o}z} et~al.,}{{Mr{\'o}z}
  et~al.}{2019}]{Mroz2019}
{Mr{\'o}z} P.,  et~al., 2019, \mn@doi [\apjs] {10.3847/1538-4365/ab426b}, \href
  {https://ui.adsabs.harvard.edu/abs/2019ApJS..244...29M} {244, 29}

\bibitem[\protect\citeauthoryear{{Mr{\'o}z}, {Udalski}, {Wyrzykowski},
  {Skowron}, {Poleski}, {Szymanski}, {Soszynski}  \& {Ulaczyk}}{{Mr{\'o}z}
  et~al.}{2021}]{Mroz2021b}
{Mr{\'o}z} P.,  {Udalski} A.,  {Wyrzykowski} L.,  {Skowron} J.,  {Poleski} R.,
  {Szymanski} M.,  {Soszynski} I.,   {Ulaczyk} K.,  2021, arXiv e-prints, \href
  {https://ui.adsabs.harvard.edu/abs/2021arXiv210713697M} {p. arXiv:2107.13697}

\bibitem[\protect\citeauthoryear{{Navarro}, {Minniti}  \& {Contreras
  Ramos}}{{Navarro} et~al.}{2017}]{Navarro2017}
{Navarro} M.~G.,  {Minniti} D.,   {Contreras Ramos} R.,  2017, \mn@doi [\apjl]
  {10.3847/2041-8213/aa9b29}, \href
  {https://ui.adsabs.harvard.edu/abs/2017ApJ...851L..13N} {851, L13}

\bibitem[\protect\citeauthoryear{{Navarro}, {Minniti}  \&
  {Contreras-Ramos}}{{Navarro} et~al.}{2018}]{Navarro2018Longitude}
{Navarro} M.~G.,  {Minniti} D.,   {Contreras-Ramos} R.,  2018, \mn@doi [\apjl]
  {10.3847/2041-8213/aae08a}, \href
  {https://ui.adsabs.harvard.edu/abs/2018ApJ...865L...5N} {865, L5}

\bibitem[\protect\citeauthoryear{{Navarro}, {Minniti}, {Pullen}  \&
  {Ramos}}{{Navarro} et~al.}{2020a}]{NavarroLatitude}
{Navarro} M.~G.,  {Minniti} D.,  {Pullen} J.,   {Ramos} R.~C.,  2020a, \mn@doi
  [\apj] {10.3847/1538-4357/ab5e4c}, \href
  {https://ui.adsabs.harvard.edu/abs/2020ApJ...889...56N} {889, 56}

\bibitem[\protect\citeauthoryear{{Navarro}, {Contreras Ramos}, {Minniti},
  {Pullen}, {Capuzzo-Dolcetta}  \& {Lucas}}{{Navarro}
  et~al.}{2020b}]{NavarroForesaken}
{Navarro} M.~G.,  {Contreras Ramos} R.,  {Minniti} D.,  {Pullen} J.,
  {Capuzzo-Dolcetta} R.,   {Lucas} P.~W.,  2020b, \mn@doi [\apj]
  {10.3847/1538-4357/ab7a9d}, \href
  {https://ui.adsabs.harvard.edu/abs/2020ApJ...893...65N} {893, 65}

\bibitem[\protect\citeauthoryear{{Navarro}, {Minniti}  \& {Contreras
  Ramos}}{{Navarro} et~al.}{2020c}]{NavarroFarDisk}
{Navarro} M.~G.,  {Minniti} D.,   {Contreras Ramos} R.,  2020c, \mn@doi [\apj]
  {10.3847/1538-4357/abaf00}, \href
  {https://ui.adsabs.harvard.edu/abs/2020ApJ...902...35N} {902, 35}

\bibitem[\protect\citeauthoryear{{Paczynski}}{{Paczynski}}{1986}]{Paczynski1986}
{Paczynski} B.,  1986, \mn@doi [\apj] {10.1086/163919}, \href
  {https://ui.adsabs.harvard.edu/abs/1986ApJ...301..503P} {301, 503}

\bibitem[\protect\citeauthoryear{{Paczynski}}{{Paczynski}}{1996}]{Paczynski1996}
{Paczynski} B.,  1996, \mn@doi [Annual Review of Astronomy and Astrophysics]
  {10.1146/annurev.astro.34.1.419}, \href
  {https://ui.adsabs.harvard.edu/abs/1996ARA&A..34..419P} {34, 419}

\bibitem[\protect\citeauthoryear{{Padmanabhan} et~al.,}{{Padmanabhan}
  et~al.}{2008}]{ubercal}
{Padmanabhan} N.,  et~al., 2008, \mn@doi [\apj] {10.1086/524677}, \href
  {https://ui.adsabs.harvard.edu/abs/2008ApJ...674.1217P} {674, 1217}

\bibitem[\protect\citeauthoryear{{Penny}, {Gaudi}, {Kerins}, {Rattenbury},
  {Mao}, {Robin}  \& {Calchi Novati}}{{Penny} et~al.}{2019}]{Penny2019}
{Penny} M.~T.,  {Gaudi} B.~S.,  {Kerins} E.,  {Rattenbury} N.~J.,  {Mao} S.,
  {Robin} A.~C.,   {Calchi Novati} S.,  2019, \mn@doi [\apjs]
  {10.3847/1538-4365/aafb69}, \href
  {https://ui.adsabs.harvard.edu/abs/2019ApJS..241....3P} {241, 3}

\bibitem[\protect\citeauthoryear{{Poindexter}, {Afonso}, {Bennett},
  {Glicenstein}, {Gould}, {Szyma{\'n}ski}  \& {Udalski}}{{Poindexter}
  et~al.}{2005}]{Po05}
{Poindexter} S.,  {Afonso} C.,  {Bennett} D.~P.,  {Glicenstein} J.-F.,  {Gould}
  A.,  {Szyma{\'n}ski} M.~K.,   {Udalski} A.,  2005, \mn@doi [\apj]
  {10.1086/468182}, \href
  {https://ui.adsabs.harvard.edu/abs/2005ApJ...633..914P} {633, 914}

\bibitem[\protect\citeauthoryear{{Refsdal}}{{Refsdal}}{1964}]{Refdal1964}
{Refsdal} S.,  1964, \mn@doi [\mnras] {10.1093/mnras/128.4.295}, \href
  {https://ui.adsabs.harvard.edu/abs/1964MNRAS.128..295R} {128, 295}

\bibitem[\protect\citeauthoryear{{Rybicki}, {Wyrzykowski}, {Klencki}, {de
  Bruijne}, {Belczy{\'n}ski}  \& {Chru{\'s}li{\'n}ska}}{{Rybicki}
  et~al.}{2018}]{Rybicki2018}
{Rybicki} K.~A.,  {Wyrzykowski} {\L}.,  {Klencki} J.,  {de Bruijne} J.,
  {Belczy{\'n}ski} K.,   {Chru{\'s}li{\'n}ska} M.,  2018, \mn@doi [\mnras]
  {10.1093/mnras/sty356}, \href
  {https://ui.adsabs.harvard.edu/abs/2018MNRAS.476.2013R} {476, 2013}

\bibitem[\protect\citeauthoryear{{Sahu} et~al.,}{{Sahu}
  et~al.}{2017}]{Sahu2017}
{Sahu} K.~C.,  et~al., 2017, \mn@doi [Science] {10.1126/science.aal2879}, \href
  {https://ui.adsabs.harvard.edu/abs/2017Sci...356.1046S} {356, 1046}

\bibitem[\protect\citeauthoryear{{Sahu} et~al.,}{{Sahu}
  et~al.}{2022}]{Sahu2022}
{Sahu} K.~C.,  et~al., 2022, arXiv e-prints, \href
  {https://ui.adsabs.harvard.edu/abs/2022arXiv220113296S} {p. arXiv:2201.13296}

\bibitem[\protect\citeauthoryear{{Saito} et~al.,}{{Saito}
  et~al.}{2013}]{Saito2013}
{Saito} R.~K.,  et~al., 2013, \mn@doi [\aap] {10.1051/0004-6361/201321260},
  \href {https://ui.adsabs.harvard.edu/abs/2013A&A...554A.123S} {554, A123}

\bibitem[\protect\citeauthoryear{{Schechter}, {Mateo}  \& {Saha}}{{Schechter}
  et~al.}{1993}]{dophot1}
{Schechter} P.~L.,  {Mateo} M.,   {Saha} A.,  1993, \mn@doi [\pasp]
  {10.1086/133316}, \href
  {https://ui.adsabs.harvard.edu/abs/1993PASP..105.1342S} {105, 1342}

\bibitem[\protect\citeauthoryear{{Shao} \& {Li}}{{Shao} \&
  {Li}}{2020}]{Shao2020}
{Shao} Y.,  {Li} X.-D.,  2020, \mn@doi [\apj] {10.3847/1538-4357/aba118}, \href
  {https://ui.adsabs.harvard.edu/abs/2020ApJ...898..143S} {898, 143}

\bibitem[\protect\citeauthoryear{Sharan}{Sharan}{2019}]{Sharan2019}
Sharan A.,  2019, Master's thesis, University of Auckland

\bibitem[\protect\citeauthoryear{{Shvartzvald}, {Bryden}, {Gould}, {Henderson},
  {Howell}  \& {Beichman}}{{Shvartzvald} et~al.}{2017}]{Shvartzvald2017}
{Shvartzvald} Y.,  {Bryden} G.,  {Gould} A.,  {Henderson} C.~B.,  {Howell}
  S.~B.,   {Beichman} C.,  2017, \mn@doi [\aj] {10.3847/1538-3881/153/2/61},
  \href {https://ui.adsabs.harvard.edu/abs/2017AJ....153...61S} {153, 61}

\bibitem[\protect\citeauthoryear{Skilling et~al.}{Skilling
  et~al.}{2006}]{Skilling2006}
Skilling J.,  et~al., 2006, Bayesian analysis, 1, 833

\bibitem[\protect\citeauthoryear{{Smith} et~al.,}{{Smith} et~al.}{2002}]{Sm02}
{Smith} M.~C.,  et~al., 2002, \mn@doi [\mnras]
  {10.1046/j.1365-8711.2002.05811.x}, \href
  {https://ui.adsabs.harvard.edu/abs/2002MNRAS.336..670S} {336, 670}

\bibitem[\protect\citeauthoryear{{Smith}, {Mao}  \& {Paczy{\'n}ski}}{{Smith}
  et~al.}{2003}]{Sm03}
{Smith} M.~C.,  {Mao} S.,   {Paczy{\'n}ski} B.,  2003, \mn@doi [\mnras]
  {10.1046/j.1365-8711.2003.06183.x}, \href
  {https://ui.adsabs.harvard.edu/abs/2003MNRAS.339..925S} {339, 925}

\bibitem[\protect\citeauthoryear{{Smith}, {Wo{\'z}niak}, {Mao}  \&
  {Sumi}}{{Smith} et~al.}{2007}]{Smith2007}
{Smith} M.~C.,  {Wo{\'z}niak} P.,  {Mao} S.,   {Sumi} T.,  2007, \mn@doi
  [\mnras] {10.1111/j.1365-2966.2007.12130.x}, \href
  {https://ui.adsabs.harvard.edu/abs/2007MNRAS.380..805S} {380, 805}

\bibitem[\protect\citeauthoryear{{Smith} et~al.,}{{Smith} et~al.}{2018}]{virac}
{Smith} L.~C.,  et~al., 2018, \mn@doi [\mnras] {10.1093/mnras/stx2789}, \href
  {https://ui.adsabs.harvard.edu/abs/2018MNRAS.474.1826S} {474, 1826}

\bibitem[\protect\citeauthoryear{{Speagle}}{{Speagle}}{2020}]{DYNESTY}
{Speagle} J.~S.,  2020, \mn@doi [\mnras] {10.1093/mnras/staa278}, \href
  {https://ui.adsabs.harvard.edu/abs/2020MNRAS.493.3132S} {493, 3132}

\bibitem[\protect\citeauthoryear{{Specht}, {Kerins}, {Awiphan}  \&
  {Robin}}{{Specht} et~al.}{2020}]{Specht2020}
{Specht} D.,  {Kerins} E.,  {Awiphan} S.,   {Robin} A.~C.,  2020, \mn@doi
  [\mnras] {10.1093/mnras/staa2375}, \href
  {https://ui.adsabs.harvard.edu/abs/2020MNRAS.498.2196S} {498, 2196}

\bibitem[\protect\citeauthoryear{Strope, Schaefer  \& Henden}{Strope
  et~al.}{2010}]{Str10}
Strope R.~J.,  Schaefer B.~E.,   Henden A.~A.,  2010, \mn@doi [The Astronomical
  Journal] {10.1088/0004-6256/140/1/34}, 140, 34–62

\bibitem[\protect\citeauthoryear{{Sutherland} et~al.,}{{Sutherland}
  et~al.}{2015}]{sutherland15}
{Sutherland} W.,  et~al., 2015, \aap, \href
  {http://adsabs.harvard.edu/abs/2015A%26A...575A..25S} {575, A25}

\bibitem[\protect\citeauthoryear{{Udalski}, {Szyma{\'n}ski}  \&
  {Szyma{\'n}ski}}{{Udalski} et~al.}{2015}]{Udalski2015}
{Udalski} A.,  {Szyma{\'n}ski} M.~K.,   {Szyma{\'n}ski} G.,  2015, \actaa,
  \href {http://adsabs.harvard.edu/abs/2015AcA....65....1U} {65, 1}

\bibitem[\protect\citeauthoryear{{WFIRST Astrometry Working Group}
  et~al.,}{{WFIRST Astrometry Working Group} et~al.}{2019}]{WFIRST2019}
{WFIRST Astrometry Working Group} et~al., 2019, \mn@doi [Journal of
  Astronomical Telescopes, Instruments, and Systems]
  {10.1117/1.JATIS.5.4.044005}, \href
  {https://ui.adsabs.harvard.edu/abs/2019JATIS...5d4005W} {5, 044005}

\bibitem[\protect\citeauthoryear{{Walker}}{{Walker}}{1995}]{Walker1995}
{Walker} M.~A.,  1995, \mn@doi [\apj] {10.1086/176367}, \href
  {https://ui.adsabs.harvard.edu/abs/1995ApJ...453...37W} {453, 37}

\bibitem[\protect\citeauthoryear{{Wyrzykowski} \& {Mandel}}{{Wyrzykowski} \&
  {Mandel}}{2020}]{Wyrzykowski2020}
{Wyrzykowski} {\L}.,  {Mandel} I.,  2020, \mn@doi [\aap]
  {10.1051/0004-6361/201935842}, \href
  {https://ui.adsabs.harvard.edu/abs/2020A&A...636A..20W} {636, A20}

\bibitem[\protect\citeauthoryear{{Wyrzykowski} et~al.,}{{Wyrzykowski}
  et~al.}{2015}]{Wyrzykowski2015}
{Wyrzykowski} {\L}.,  et~al., 2015, \mn@doi [\apjs]
  {10.1088/0067-0049/216/1/12}, \href
  {https://ui.adsabs.harvard.edu/abs/2015ApJS..216...12W} {216, 12}

\bibitem[\protect\citeauthoryear{{Wyrzykowski} et~al.,}{{Wyrzykowski}
  et~al.}{2016}]{Wyrzykowski2016}
{Wyrzykowski} {\L}.,  et~al., 2016, \mn@doi [\mnras] {10.1093/mnras/stw426},
  \href {https://ui.adsabs.harvard.edu/abs/2016MNRAS.458.3012W} {458, 3012}

\bibitem[\protect\citeauthoryear{{Zorec} \& {Briot}}{{Zorec} \&
  {Briot}}{1997}]{Zo97}
{Zorec} J.,  {Briot} D.,  1997, \aap, \href
  {https://ui.adsabs.harvard.edu/abs/1997A&A...318..443Z} {318, 443}

\bibitem[\protect\citeauthoryear{{Zurlo} et~al.,}{{Zurlo}
  et~al.}{2018}]{Zurlo2018}
{Zurlo} A.,  et~al., 2018, \mn@doi [\mnras] {10.1093/mnras/sty1805}, \href
  {https://ui.adsabs.harvard.edu/abs/2018MNRAS.480..236Z} {480, 236}

\makeatother
\end{thebibliography}

%%%%%%%%%%%%%%%%%%%%%%%%%%%%%%%%%%%%%%%%%%%%%%%%%%

%%%%%%%%%%%%%%%%% APPENDICES %%%%%%%%%%%%%%%%%%%%%

\appendix
\section{Photometry and astrometry tables}
This sections contains the sample photometry and astrometry data table for the dark lens events. The full table is available in the online material associated with this paper. 
\begingroup
\renewcommand{\arraystretch}{1.4}
\begin{table*}
\centering
\begin{tabular}{llcccccc} 
\hline
Event ID & $t_{\text{obs}}$ [MJD] & $K_s$ [mag] & $K_s$ error [mag] & RA [deg] & RA error [mas] & DEC [deg] & DEC error [mas] \\
 \hline
VVV-2013-DSC-0437 & 55284.39514979 & 13.993 & 0.017 & 259.69282535 & 7.21 & -39.38394611 & 7.36 \\  
VVV-2013-DSC-0437 & 55284.39597758 & 13.990 & 0.009 & 259.69282757 & 4.16 & -39.38394671 & 4.66 \\ 
VVV-2013-DSC-0437 & 55309.34561593 & 13.993 & 0.019 & 259.69282576 & 7.87 & -39.38394630 & 9.44 \\  
VVV-2013-DSC-0437 & 55428.15944366 & 13.970 & 0.020 & 259.69282898 & 11.30 & -39.38394635 & 10.95 \\
VVV-2013-DSC-0437 & 55428.16382880 & 13.974 & 0.019 & 259.69282095 & 11.56 & -39.38394895 & 8.76 \\ 
... & ... & ... & ... & ... & ... & ... & ... \\
\hline
 
\end{tabular}
\caption{Sample table containing the first 5 rows of VVV photometry and astrometry for selected candidates. The full table is available in the online supplementary material.}
\label{table:samplephotometry}
\end{table*}
\endgroup

%%%%%%%%%%%%%%%%%%%%%%%%%%%%%%%%%%%%%%%%%%%%%%%%%%

% Don't change these lines
\bsp	% typesetting comment
\label{lastpage}
\end{document}